\documentclass[preprint]{ptephy_v1}%%%%%% to generate preprint number

\preprintnumber{XXXX-XXXX} %%% %%% Insert preprint number here

\usepackage{graphics}
\usepackage{verbatim}
\usepackage{amsmath}
\usepackage{subfig}
\usepackage{url}

\usepackage{ulem}

\graphicspath{{Figures/}}

\begin{document}

\title{Characterization techniques for fixed-field alternating gradient accelerators and beam studies using the KURRI 150 MeV proton FFAG}

\author{S. L. Sheehy}
\affil{John Adams Institute, University of Oxford, Keble Rd, Oxford, OX1 3RH, UK \email{suzie.sheehy@physics.ox.ac.uk}}
\author{D. J. Kelliher}
\author{S. Machida}
\author{C. Rogers}
\author{C. R. Prior}
\affil{ASTeC, STFC Rutherford Appleton Laboratory, Harwell Oxford, Didcot, OX110QX, United Kingdom}
\author{L. Volat}
\affil{Grenoble Institute of Technology, Phelma, Grenoble, France\\}
\author{M. Haj Tahar}
\affil{Brookhaven National Laboratory, Long Island, Upton, USA}
\author{Y. Ishi}
\author{Y. Kuriyama}
\author{M. Sakamoto}
\author{T. Uesugi}
\author{Y. Mori}
\affil{Kyoto University Research Reactor Institute, Kumatori, Osaka, 590-0494, Japan}

\begin{abstract}In this paper we describe the methods and tools used to characterize a 150 MeV proton scaling Fixed Field Alternating Gradient (FFAG) accelerator at Kyoto University Research Reactor Institute. Many of the techniques used are unique to this class of machine and are thus of relevance to any future FFAG accelerator. For the first time we detail systematic studies undertaken to improve the beam quality of the FFAG. The control of beam quality in this manner is crucial to demonstrating high power operation of FFAG accelerators in future.
\end{abstract}

\subjectindex{G07}

\maketitle

\section{Introduction \label{Section1}}

The first Fixed Field Alternating Gradient (FFAG) accelerators were constructed in the 1950's and 60's as part of the MURA project~\cite{Symon2003}. However, only electron machines were constructed at the time, no hadron machines. Interest in this type of accelerator for hadrons has been revitalized since the late 1990's with technological developments such as low Q rf cavities and precise 3D modeling of magnets. As a result, a proof of principle proton FFAG accelerator with a high repetition rate ($\approx1$\,kHz) was built at KEK~\cite{aiba00} and a prototype model for medical applications with protons up to 150 MeV in kinetic energy~\cite{Adachi2001} was also successfully commissioned.

FFAG accelerators could sustain a high average current thanks to their use of DC magnets, which allows an increase in the repetition rate limited only by the rf system. This could result in a higher average current than a synchrotron while operating with a similar or lower bunch charge. Operating an FFAG with a similar bunch charge and space charge tune shift as a synchrotron would produce a much greater beam power than existing machines.

In 2012 beam experiments using the FFAG proton driver for the ADSR system at KURRI were proposed to demonstrate high beam power capability in FFAGs and subsequently an international collaboration was set up. Three weeks of dedicated beam time for characterization purposes were allocated in March 2014 with a further week in June 2014. 

FFAG accelerators are frequently described as a cross between a synchrotron and a cyclotron. In fact, their dynamics are a mixture of the two. The temporally fixed magnetic field means that the beam moves outward radially during acceleration and it is not possible to define a reference orbit in the real machine. At the same time strong focusing ensures that beams can remain stable up to high energies, but the magnetic field required to stabilize the betatron tunes over the energy range must follow a precise function of radius. In the non-relativistic regime the revolution time can vary considerably, so these machines are usually far from being isochronous. This combination of a lack of reference orbit and a varying revolution time presents new challenges in both the simulation and experimental characterization of these machines.

In simulations, for instance, we cannot assume an ideal closed orbit independent of the beam momentum. As such, the reference coordinate system cannot be specified with respect to the physical lattice magnets, for instance the central axis of quadrupole magnets, as would be assumed in a synchrotron. In the experimental case, FFAGs require rf cavities with a wide physical aperture in the horizontal direction. Any magnetic core material such as ferrite or magnetic alloy used to tune the rf frequency is required to have a wide aperture as well~\cite{Yonemura2007}. 

However, there are several advantages to this setup, principally that the orbit position can be controlled by the rf frequency. Once we fix the rf frequency, the synchronous momentum and thus the beam position as a function of radius are determined. This turns out to be a useful property to characterize FFAGs and this concept will be utilized in several later parts of this work.  

In this paper we describe systematic studies performed to understand the accelerator and improve the beam quality of the FFAG. We present methods for determining the basic lattice and beam parameters such as the momentum compaction factor, or field index $k$ which determines the magnetic field variation with radius, the closed orbit and correction of closed orbit distortion. We also perform measurements of dispersion and orbit matching, as well as measurements of the betatron tunes and present a method to determine the beam energy lost on the thin carbon stripping foil.

\section{The 150 MeV KURRI ADSR-FFAG}

At Kyoto University Research Reactor Institute (KURRI), a 150 MeV FFAG accelerator similar in design to the KEK 150 MeV prototype of medical accelerators was adopted as a proton driver for an Accelerator Driven Sub-critical Reactor (ADSR) system and delivered the first beam to reactor users in 2009~\cite{PYEON2009, ishi10}. This FFAG accelerator initially used a two-stage FFAG booster system as an injector, but recently it has been upgraded to use a H\textsuperscript{-} linac to inject beam using charge exchange injection to increase the beam intensity~\cite{mori11}. 

\begin{figure}[htbp]
\begin{center}
\includegraphics[width=0.6\linewidth]{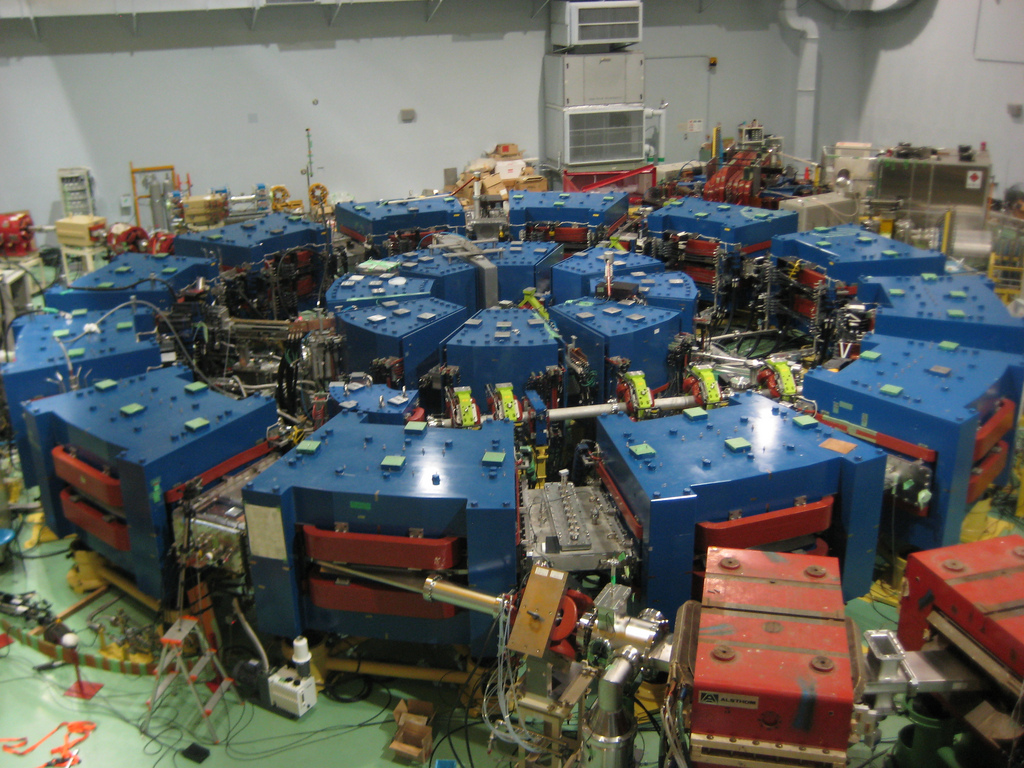}
\caption{150\,MeV ADSR-FFAG at KURRI (the larger of the two rings shown). The extraction line shown in the foreground demonstrates that the beam is extracted from the side of a ``return-yoke free" magnet. Injection from the H\textsuperscript{-} linac occurs in the top left of the image.}
\label{figure1}
\end{center}
\end{figure}

The ADSR-FFAG accelerator is shown in Fig.~\ref{figure1} and its main parameters are given in Table~\ref{table1}. When it was constructed this machine incorporated a number of key technological innovations. The first was the development of broadband low Q rf cavities using Magnetic Alloy (MA) material~\cite{Tanigaki2005} which were first used in the PoP (Proof of Principle) FFAG~\cite{aiba00}. The second development was the design of a ``return-yoke free" magnet for the 150\,MeV FFAG~\cite{Adachi2001}. In this development the magnetic flux in the defocusing-focusing-defocusing (DFD) triplet is mainly created in the focusing (F) magnet, extends through the F magnet gap and is returned through the gap of the defocusing (D) magnet. At each corner of the magnet there is a `shunt yoke' to allow some magnetic flux return and to provide mechanical stability. The key feature of these magnets is that a beam can be injected or extracted from the side of them. These unique magnets can be seen in the ADSR-FFAG in Fig.~\ref{figure1}.

\begin{table}[htbp]
\caption{Constants and experimental parameters}
\label{table1}
\centering
\begin{tabular}{|l||c||c|}
\hline
$R_{0}$, reference radius & 4.54 & m\\
Focusing structure & DFD & \\
$N_{cells}$ & 12 & \\
Injection Energy & 11 & MeV\\
Extraction Energy & 100 to 150 & MeV\\
$k$, field index & 7.6 &  \\
$f_{rf}$ & 1.6-5.2 & MHz\\
$V_{rf}$ & $\leq 4.0$ & kV\\
$B_{max}$ & 1.6 & T\\
\hline
\end{tabular}
\end{table}

At present, the number of protons per bunch is around $3\times 10^{9}$ which corresponds to a 10\,nA average current, injected from the linac which is usually operated at 20\,Hz. At extraction the bunch length is around 0.1\,$\mu$s. The only corrector magnets present in the ring are dipole corrector poles located above and below the flange of the main rf cavity.

\subsection{Beam optics and operation}

The ADSR-FFAG follows the scaling principle~\cite{Kolomensky1966, Symon1956}. This is satisfied when the median plane field has the magnetic field profile of Eq.~(\ref{eqn:field}) at any azimuth, where $r$ is radius measured from the machine centre and $B_{y,0}(\theta)$ is the vertical field at a reference radius $r_{0}$ and $k$ is the mean field index.

\begin{equation}
B_{y}(r, \theta)=B_{y,0}(\theta)\left(\frac{r}{r_{0}}\right)^{k}
\label{eqn:field}
\end{equation}

The magnetic field profile is configured so that the orbits for different momenta are similar; each orbit is simply enlarged in radius while maintaining the same shape. As a result, the transverse tune should be constant. In reality, however, imperfection of the magnets and interference with individual elements in the lattice break the symmetry and the scaling principle. This will be discussed in more detail later.

The H\textsuperscript{-} linac is used as an injector to deliver beams up to 11\,MeV. Employing H\textsuperscript{-} charge exchange injection enables the accumulation of more particles without increasing the beam emittance. This is particularly relevant for one main item of future study, that is the study of the space charge limit in the FFAG without beam loss.

\begin{figure}[h!]
\begin{center}
\includegraphics[width=0.6\linewidth]{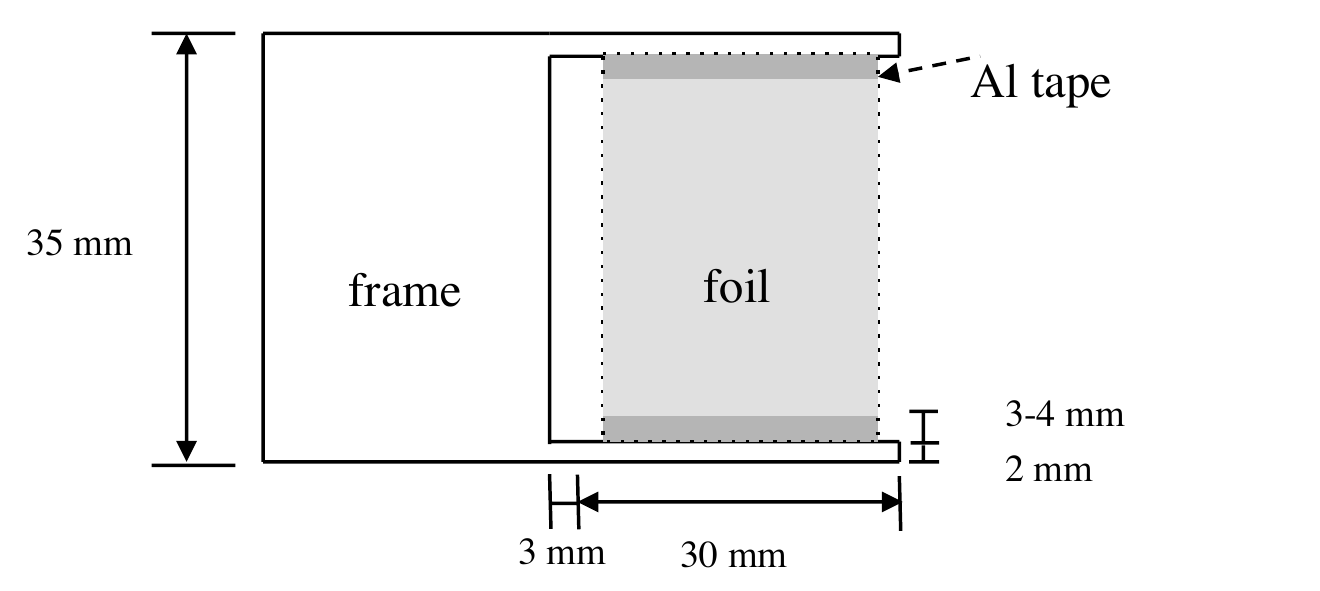}
\caption{Carbon foil and frame layout used for charge exchange injection.}
\label{foil}
\end{center}
\end{figure}

A carbon foil with a design thickness of 20 $\mu g/\textrm{cm}^2$ is mounted on a holder as shown in Fig.~\ref{foil} and inserted in the centre of a focusing magnet for charge exchange injection. The radial position of the foil is adjustable. There is no orbit bump with pulsed magnets as the beam simply moves off the foil radially during acceleration. Multiple scattering and energy loss at the foil with an 11 MeV proton beam is not negligible and measuring these effects is one of the on-going experimental efforts of the collaboration. 

Beam extraction is not required to perform the present experiments, however the beam is usually extracted within a single turn by a dual kicker and septum system.

\subsection{Diagnostics}
\begin{figure}[h!]
\begin{center}
\includegraphics[width=0.7\linewidth]{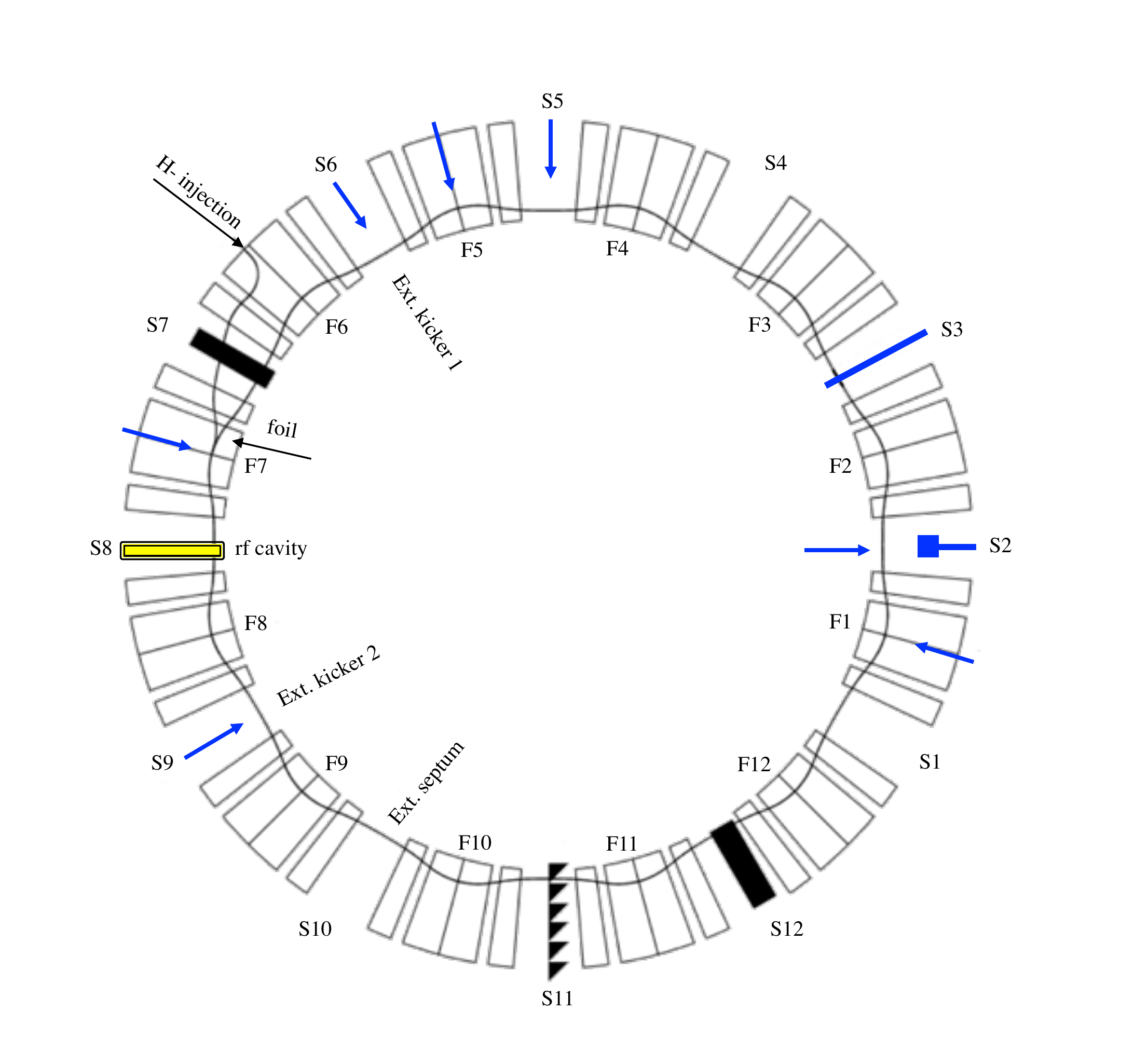}
\caption{Layout of the ADSR-FFAG diagnostics. Arrows correspond to the ports available for radial probes. A vertical perturbator is located in S3 and a horizontal perturbator in S2. At S12 a vertical beam position monitor is located. S7 and S11 are a single plate bunch charge monitor and a triangle shaped electrode, respectively. The beam is injected into F6 where a Faraday cup or phosphor screen can be located, and the stripping foil is located in F7. }
\label{diagnosticlayout}
\end{center}
\end{figure}

The layout of the FFAG and location of the diagnostics referred to throughout this work are shown in Fig.~\ref{diagnosticlayout}. In this section we give a brief overview of each type of diagnostic or beam manipulating device.

\paragraph{Radial probes}

\begin{figure}[h!]
\begin{center}
\includegraphics[width=0.5\linewidth]{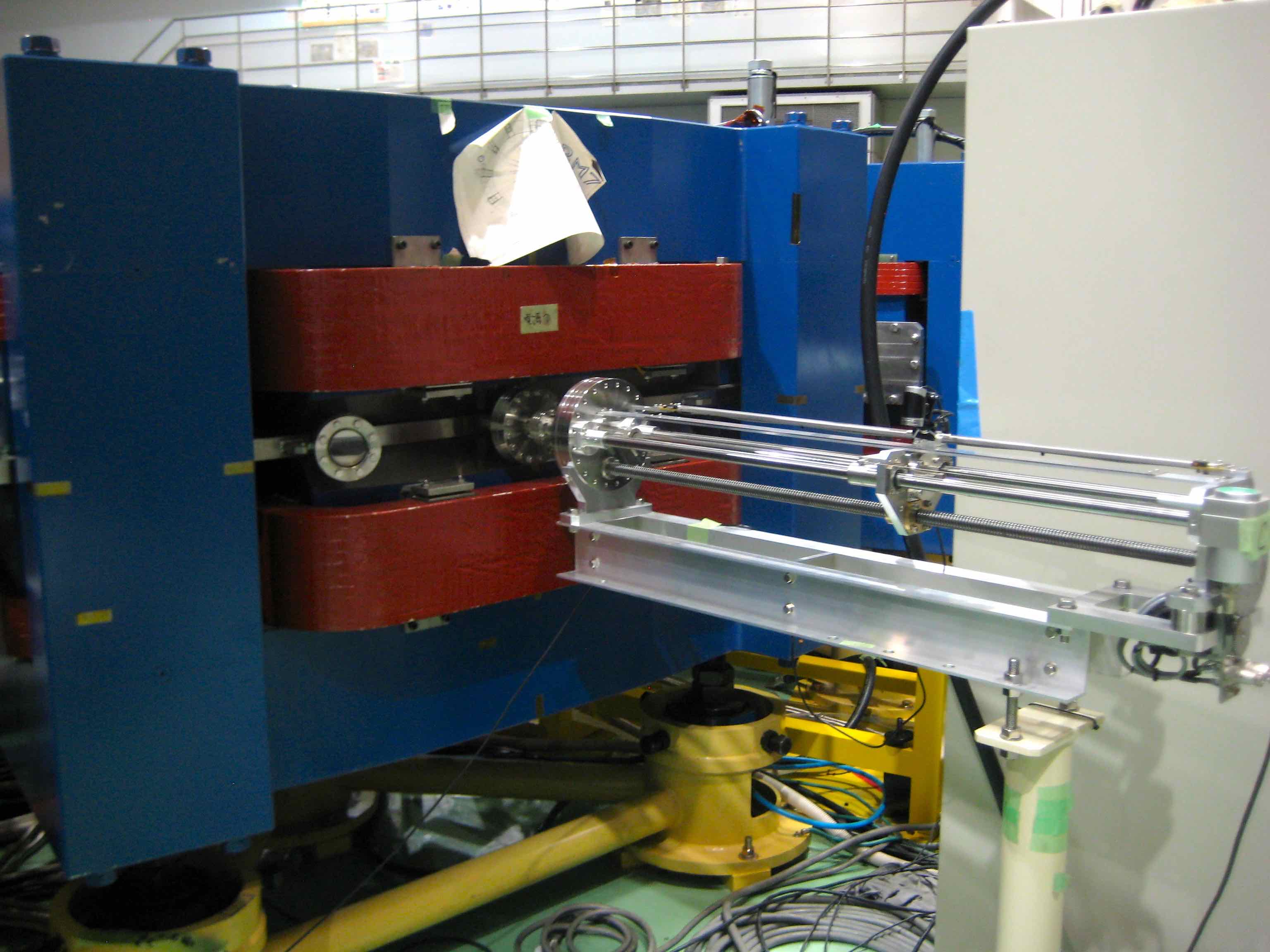}
\caption{Radial probe on stepping motor system, located in the middle of an F magnet.}
\label{radialprobe}
\end{center}
\end{figure}

At various points around the ring shown in Fig.~\ref{diagnosticlayout} there are ports for the installation of radial probes. In this machine the probes used are simply steel rods of roughly 15\,mm diameter, mounted on a stepping motor with calibrated position reading. The probes can be moved in or out of the ring radially from the control room (Fig.~\ref{radialprobe}). The radial probes are not used to read out beam current or to observe scattering or secondary emissions. In future it may be possible to install radial probes which provide beam current and current density measurements like those commonly used in cyclotrons~\cite{Dolling2010}.

\paragraph{Faraday cup}

There is a large Faraday cup which covers the aperture of the machine after injection, which has previously been used to ensure good transmission of the beam from the transport line into the ring. The main magnetic field is used to deflect secondary electrons. This device is not utilized in present experiments.

\paragraph{Beam position and bunch charge monitors}

\begin{figure}[htbp]
\begin{center}
\includegraphics[width=0.7\linewidth]{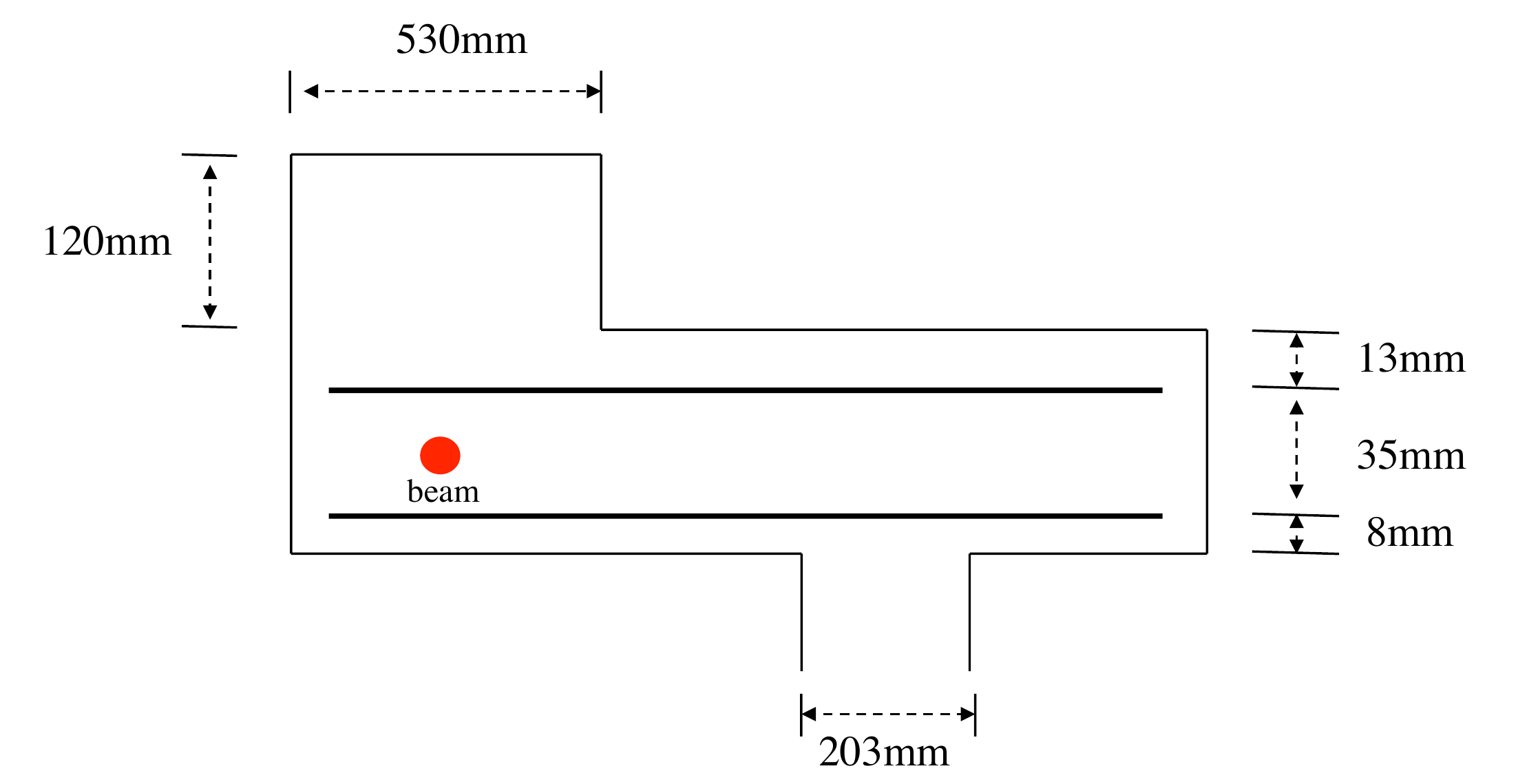}
\caption{Layout of the vertical BPM including a cross section of the vacuum chamber surrounding the double plate electrodes.}
\label{bunchmonitor}
\end{center}
\end{figure}

The main vertical beam position monitor (BPM) consists of two electrodes which span the radial extent of the vacuum chamber, which is roughly 90\,cm across, shown in Fig.~\ref{bunchmonitor}. This allows a non-destructive method of measuring vertical coherent oscillations and when calibrated can also provide a vertical position measurement. It is located in S12 approximately half way around the ring from the injection point. There is in addition a single plate electrode in the injection region in S7 that can monitor vertical oscillations but cannot provide a vertical position measurement without a second electrode. The beam position monitors are also used as a bunch charge monitors when the signal of the two electrodes are added. This enables a relative turn-by-turn measurement of bunch charge which provides information about beam loss.

In S11 there are also six triangle shaped electrodes. These are of limited use in the current configuration for a position monitor as it is not possible to differentiate between horizontal or vertical oscillations without a second set of triangle electrodes. However, the triangle electrode monitor has been used in some tune measurements to observe transverse coherent oscillations, in which the horizontal and vertical oscillation frequencies are easy to distinguish. Both the single plate electrode and the triangle plate may be used to cross-check measurements in the event of unexpected results or high levels of noise on the double plate electrode monitor.

\paragraph{Movable triangle BPM}

\begin{figure}[h!]
\begin{center}
\includegraphics[width=0.4\linewidth]{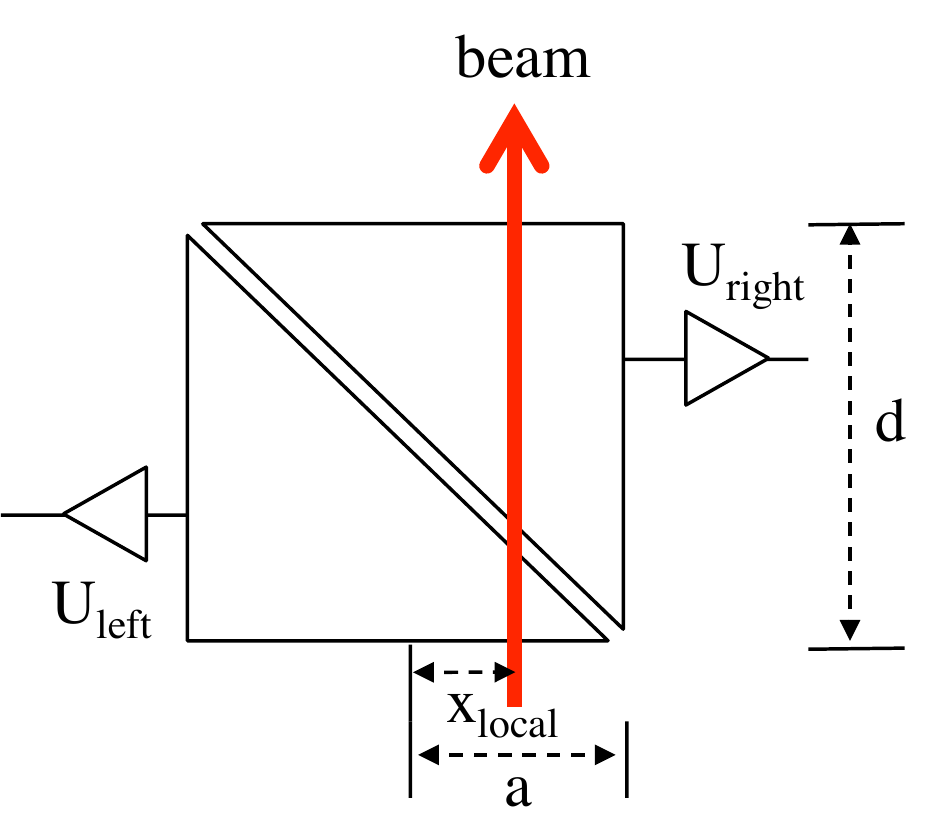}
\caption{Triangle plate monitor can be adjusted radially to provide horizontal beam position measurements over the full aperture.}
\label{triangleplate}
\end{center}
\end{figure}

To enable horizontal coherent oscillation and position measurements over a large radial range, a movable triangle BPM was developed. This consists of two sets of triangle plates located above and below the beam position, mounted on a radial mover. The BPM is stationary during a measurement and is able to measure the beam position provided the beam is located within its aperture. The key advantage of this device is that can be relocated to measure a different section of the radial aperture without breaking vacuum. The layout of each set of triangles is similar to a linear-cut BPM~\cite{Forck2009} as shown in Fig.~\ref{triangleplate}. While it was not used for position measurements in this paper, the position can in principle be calculated with respect to the induced voltages as 
\begin{equation}
x=x_{\textrm{local}}+x_{\textrm{off}}=a.\frac{U_{\textrm{right}}-U_{\textrm{left}}}{U_{\textrm{right}}+U_{\textrm{left}}}+x_{\textrm{off}},
\end{equation}

where a is the half aperture width of the triangle plate and $x_{\textrm{off}}$ is the position offset determined by the radial mover. The width of the device is $d=10$\,cm.

\paragraph{Horizontal and vertical perturbators}

\begin{figure}[h!]
\begin{center}
\includegraphics[width=0.4\linewidth]{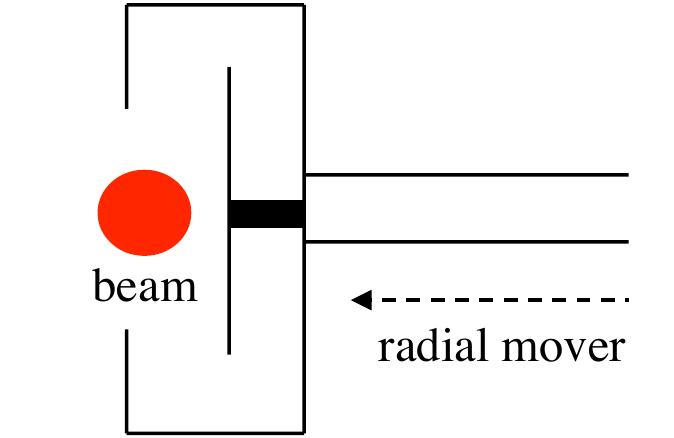}
\caption{Horizontal perturbator can be moved radially to drive horizontal coherent oscillations for tune measurements.}
\label{perturbator}
\end{center}
\end{figure}

It is sometimes necessary to introduce a known perturbation to the circulating beam, for example during betatron tune measurements in Section~\ref{Section5}. For this purpose there is a radially movable rf shaker or `perturbator' to drive coherent oscillations in the horizontal direction. This consists of a single plate driven with a sinusoidal rf signal, with a surrounding C-shaped shielding to localize the perturbation in the vicinity of the beam. The frequency of the driving signal is swept until a response is seen in the oscillations of the beam. The layout of the device can be seen in Fig.~\ref{perturbator}. For vertical perturbations a single plate, similar in layout to the vertical BPM, is driven with a sinusoidal rf signal in a similar manner.

\section{Lattice and beam parameters with accelerated orbits}
\label{Section3}
Unlike cyclotrons, FFAG accelerators usually do not satisfy the isochronous condition (for a counter example see~\cite{Brooks2013}). At each momentum, $p$, the beam has a different revolution frequency, $f$, as well as a different circumference, $C$. The slippage factor $\eta$ is defined as,

\begin{equation}
\eta=\frac{df/f}{dp/p}=\frac{1}{\gamma^2}-\alpha_p,
\end{equation}
where $\gamma$ is the relativistic factor and $\alpha_p$ is the momentum compaction factor defined as
\begin{equation}
\alpha_p=\frac{dC/C}{dp/p}=\frac{dR/R}{dp/p}.
\label{Eqn4}
\end{equation}

\noindent where $R$ is the equivalent radius defined as $R=C/2\pi$. 

In operating terms, this means we can control the beam momentum and orbit position in the radial direction with the rf frequency once the beam is captured and bunched in the rf bucket.

In scaling FFAGs, a parameter $k$ is defined as the measure of the momentum compaction,
\begin{equation}
k=R\frac{d\ln p}{dR}-1
\label{Eqn5}
\end{equation}
which is constant in the ideal design. In terms of the mean magnetic field $\overline{B}=p/eR$, we can write $k$ also as a mean field index,
\begin{equation}
k=\frac{R}{\overline{B}}\frac{d\bar{B}}{dR}.
\label{Eqn6}
\end{equation}

Measuring the orbit position as a function of the rf frequency gives us essential information about the lattice and the beam. When we measure the orbit position at different azimuthal points around the ring, but at the same point in the rf cycle (with the same rf frequency) it tells us how the closed orbit is distorted. This method can be repeated for a range of different rf frequencies and therefore with a range of different momenta. In addition, if we include the excitation of the dipole corrector magnet during this measurement, we can test the influence of the correction scheme.

If we measure the gradient of the orbit position with respect to the beam momentum, we can determine the momentum compaction factor and therefore $k$ value. It also gives the average dispersion function defined as

\begin{equation}
D=\frac{dR}{dp/p}=\frac{R}{k+1}.
\end{equation}

To make such a measurement, the beam position is measured by a radial probe which acts as a scraper. When the rf frequency increases and the beam moves outward, the beam is lost on the scraper and we see the beam intensity drop off at this point. This beam loss curve with respect to the rf frequency gives us additional information about the horizontal beam size. In the following section, we will discuss the details of the measurement technique and results.

\subsection{Momentum compaction factor}
\label{sec:3.1}

There are three locations at the centre of focusing magnets in the ring where radial probes, which act as scrapers, can be inserted from the outer wall port, F1, F5 and F7. The position of the probe edge is set either using a pulsed motor or manually. The probe position is monitored remotely in the control room. The circulating beam current is measured by the S7 bunch charge monitor. 

\begin{figure}[h!]
\begin{center}
\includegraphics[width=0.7\linewidth]{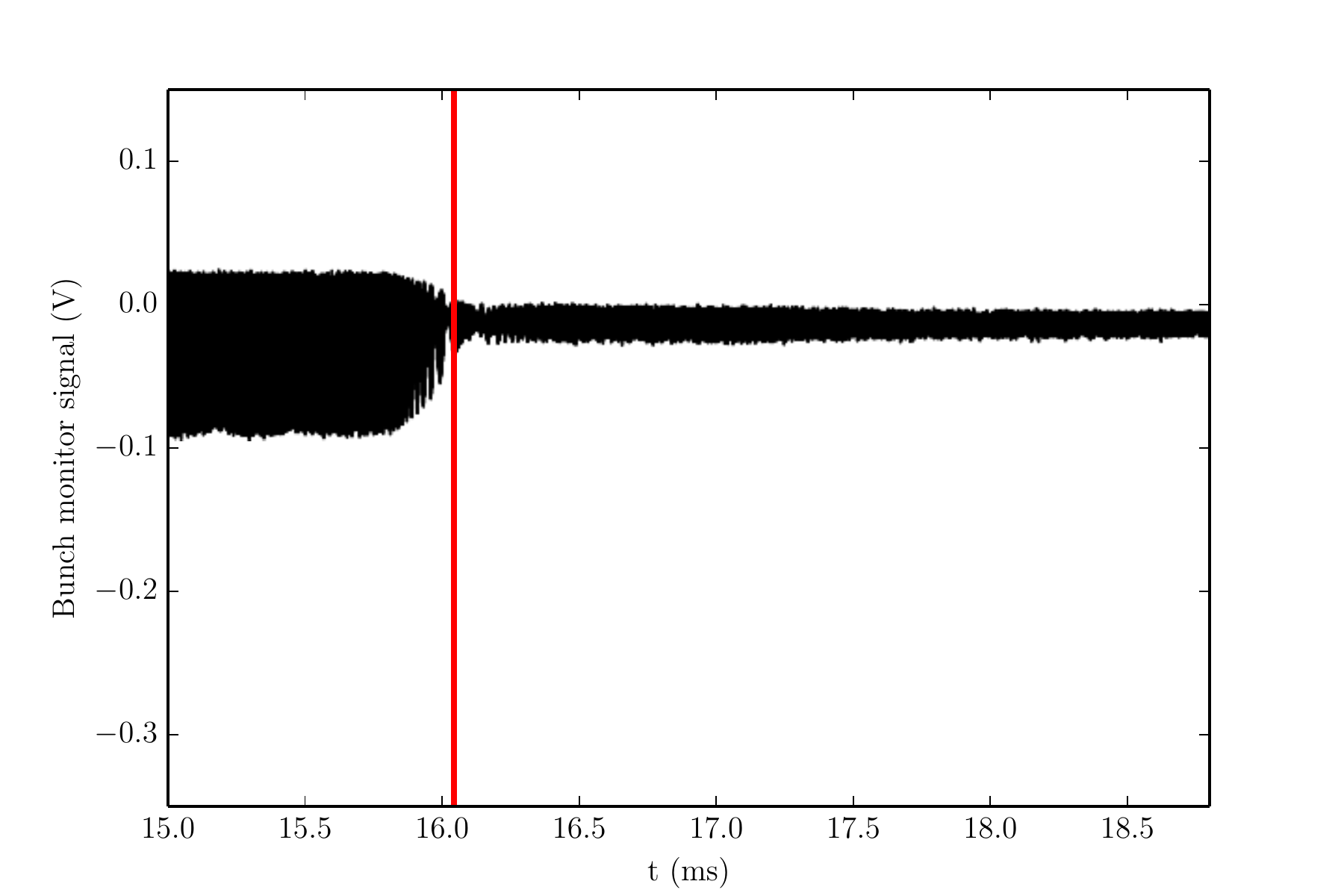}
\caption{Example of time to loss measurement when the radial position of the probe is 5.17\,m. The raw bunch charge monitor signal is shown in black. The red vertical line shows the beam loss time as found by the algorithm described in the text. The thickness of the line indicates the width of the $4 \mu$s time window in which the signal-to-noise ratio is calculated.}
\label{fig:timetoloss}
\end{center}
\end{figure}

When the beam is accelerated, the orbit moves radially until the beam hits the probe and eventually the whole beam is lost. Figure~\ref{fig:timetoloss} shows the case where the whole beam is lost at around 16\,ms upon interception with a single probe. The point of complete beam loss is defined by the time at which the signal-to-noise ratio drops below a threshold. The threshold is determined as follows.

First, the noise level $\sigma_{noise}$ is found by calculating the standard deviation within a time window at the very end of the data when it is clear there is no beam. The time window is chosen to be of much shorter duration than the fall-off time of the bunch charge monitor signal.  The time window is then progressively shifted back and the standard deviation calculated at each step until $\sigma$ reaches a multiple of $\sigma_{noise}$. In all cases studied, the criterion $\sigma \geq 2 \sigma_{noise}$ provides a reasonable measure of the loss time.

By repeating the same measurement for each radial position, we obtain the beam position as a function of time $r(t)$. The applied rf frequency pattern $f(t)$ is given and derivatives of both quantities are available so that the $k$ value can be expressed as

\begin{equation}
\frac{k(t)+1}{\gamma(t)^2}=\frac{\frac{1}{f(t)}\frac{df(t)}{dt}}{\frac{1}{r(t)}\frac{dr(t)}{dt}}+1
\label{eqn:kgamma}
\end{equation}
where we replace the average radius $dR(t)/R(t)$ by the local radius $dr(t)/r(t)$ and $\gamma(t)$ is the usual relativistic Lorentz factor.

The right hand side contains measured quantities, but $\gamma(t)$ on the left hand side is not known \textit{a priori} and must be evaluated, along with $k(t)$, consistently. Equation~(\ref{Eqn5}) can be written in terms of measured quantities as

\begin{equation}
\frac{1}{p(t)}\frac{dp(t)}{dt}=(k(t)+1)\frac{1}{r(t)}\frac{dr(t)}{dt}.
\end{equation}

\noindent Integration of both sides gives,
\begin{equation}
\int_{p(t_0)}^{p(t)}\frac{dp(t)}{p(t)}=\int_{t_0}^{t}(k(t)+1)\frac{1}{r(t)}\frac{dr(t)}{dt}dt
\end{equation}

\noindent Therefore,
\begin{equation}
p(t)=p(t_0)\exp ( \int_{t_0}^{t}(k(t)+1)\frac{1}{r(t)}\frac{dr(t)}{dt}dt).
\label{eqn:pintegral}
\end{equation}

\noindent Once we know $p(t)$, $\gamma(t)$ can be derived. In practice, Eqs.~(\ref{eqn:kgamma}) and (\ref{eqn:pintegral}) are solved iteratively to obtain $p(t)$, $\gamma(t)$ and $k(t)$. 

\begin{figure*}[!h]
\centerline{\subfloat[]{\includegraphics[width=0.7\linewidth]{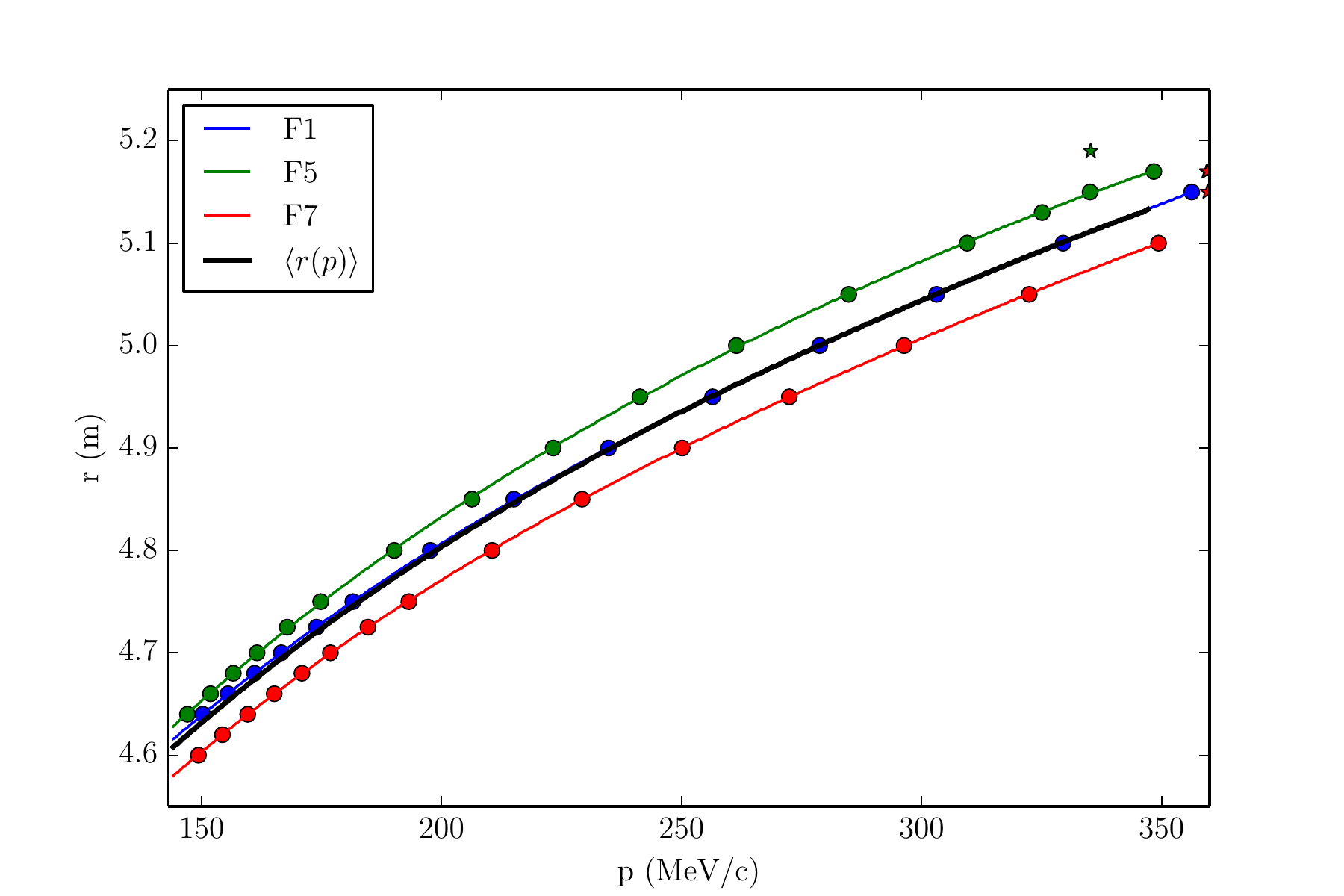}
\label{fig:r_vs_p}}}
\hfil
\centerline{\subfloat[]{\includegraphics[width=0.7\linewidth]{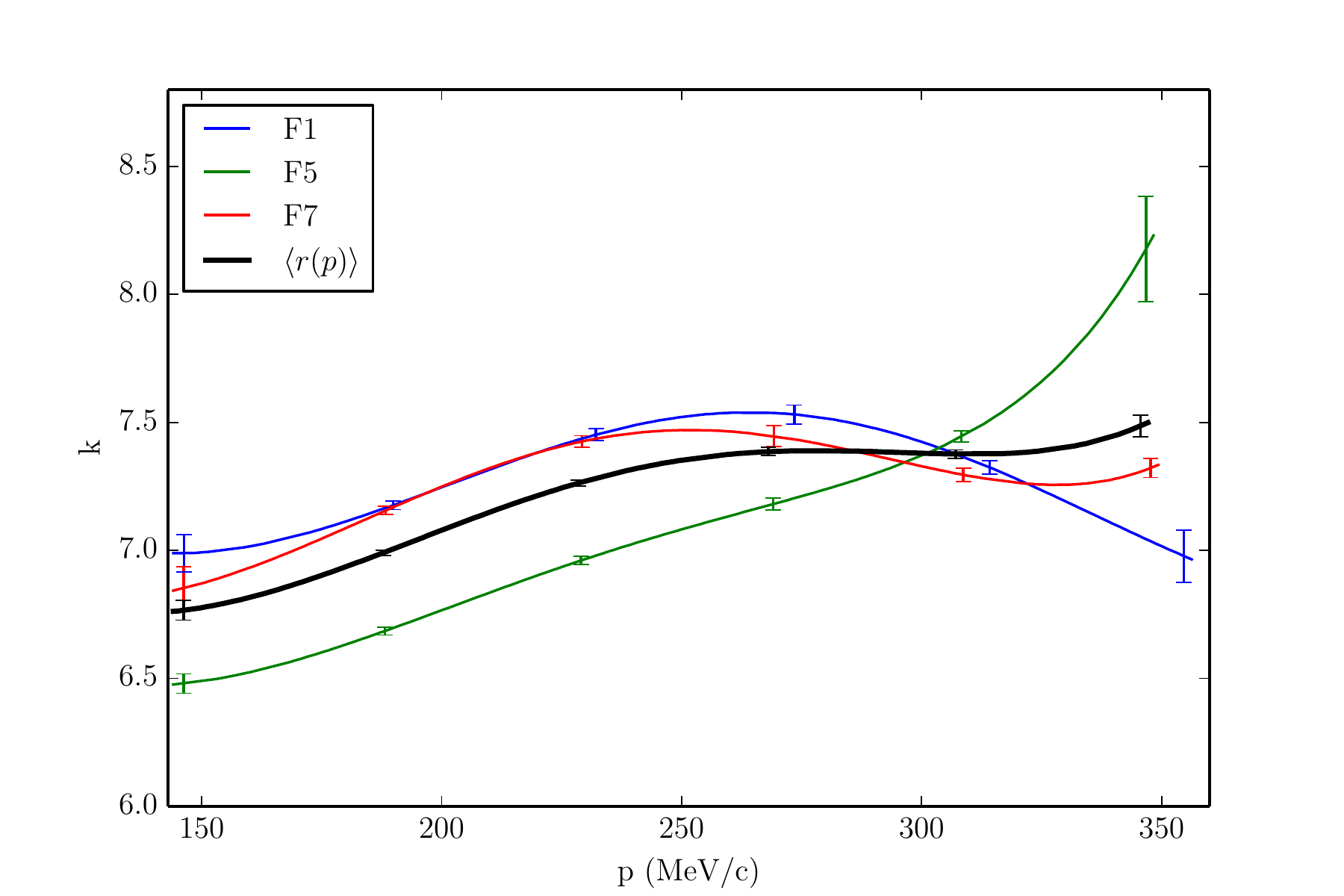}
\label{fig:k_vs_p}}}
\caption{(a): Radial probe measurements (points) and polynomial fit (lines) at the three different probe positions with respect to momentum during the acceleration cycle. The starred points are considered to be outliers and are excluded from the fit. The mean radius (black line) is found by taking the average of the three fits. (b): Calculated $k$ value at the three different probe positions with respect to momentum during the acceleration cycle. The $k$ value calculated from the mean radius fit is shown in black. The error bars are found by propagating the uncertainties in the radial probe measurements through the analysis and indicate one standard deviation in the results.}
\label{fig:r_k_vs_p}
\end{figure*}

As can be seen in Eq.~(\ref{eqn:pintegral}), the algorithm requires the initial momentum $p(t_0)$ over the time range of the integration as input. By fitting the $r(t)$ data with an appropriate function and extrapolating back in time, the injection momentum (which is measured separately) can be used. At each discrete time point $t=t_j$, the value of $k(t_j)$ that minimizes the difference in the momenta obtained from Eqs.~(\ref{eqn:kgamma}) and (\ref{eqn:pintegral}) is found using a standard root finding technique. Note, the latter equation contains in the integrand all the $k_j$ values that precede a particular time point. Finally, the $k$ value and momenta are obtained as a function of time
so that the local radius and the $k$ value can be plotted as a function of momentum as shown in Fig.~\ref{fig:r_k_vs_p}.

It is clear in the results in Fig.~\ref{fig:r_k_vs_p} the three different probes used for this measurement produce different results, particularly for the case of the F5 probe. The closed orbit distortion (COD) itself should not produce a different $k$ value at different locations as long as the shape of the distorted closed orbit is independent of momentum. The different curves of $k$ indicate that this is not the case and that the distorted closed orbit depends on the momentum. We will discuss this further in Section~\ref{section7}.

\begin{figure}[!h]
\begin{center}
\includegraphics[width=0.7\linewidth]{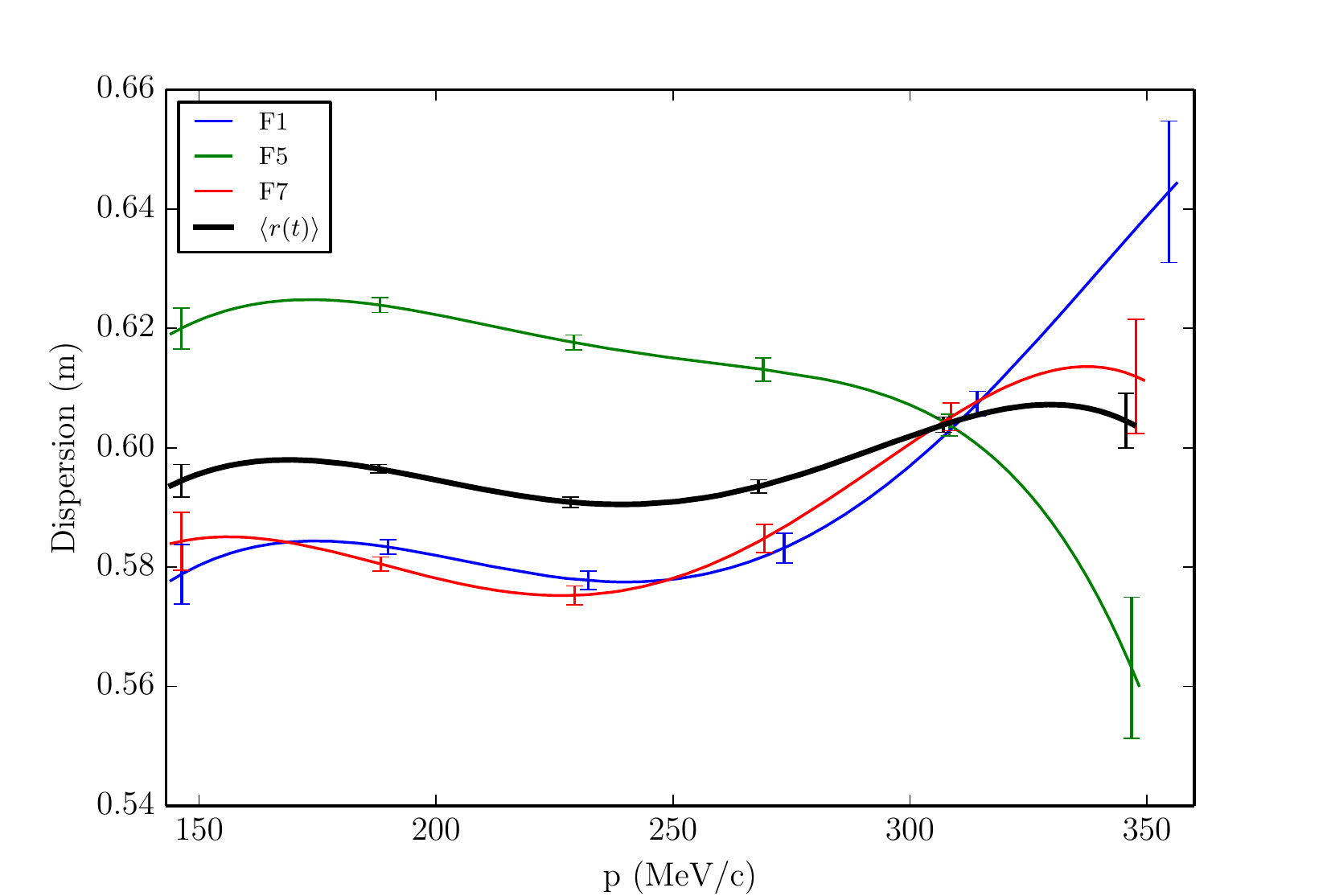}
\caption{Calculated dispersion in the ring using three different probes with respect to momentum during acceleration. The dispersion calculated from the mean radius fit is shown in black. The error bars are found by propagating the uncertainties in the radius measurements through the analysis and indicate one standard deviation in the results.}
\label{fig:dispersion}
\end{center}
\end{figure}

The dispersion function at the location of the probes is defined as
\begin{equation}
D_{p,i}=\frac{dr_{i}}{dp/p}=\frac{r_{i}}{k_{i}+1},
\end{equation}
where we have introduced the index $i$ to denote a local variable at the location $F_{i}$. Different $k$ means there is a different local dispersion function as shown in Fig.~\ref{fig:dispersion}.

\subsection{Closed orbit distortion and its correction}

Although the three probes are each inserted at the centre of focusing magnets and the local orbit radius should be the same due to the symmetry of the lattice, the measured $r(p)$ depends on the location. This indicates that the FFAG has an asymmetry owing to closed orbit distortion.

In fact, one consequence of the return-yoke free magnet design is the existence of substantial stray fields or fringe fields at the sides of the magnet. As long as the 12-fold symmetry of the field is preserved, this does not produce COD. However, the frequency tuning material in the rf cavity, MA, has a high permeability and the existence of stray field in the straight section where the rf cavity is located produces a strong interference. As a result, our hypothesis is that this is the single most dominant COD source (there is only one rf cavity in the ring).

\begin{figure}[h!]
\begin{center}
\includegraphics[width=0.7\linewidth]{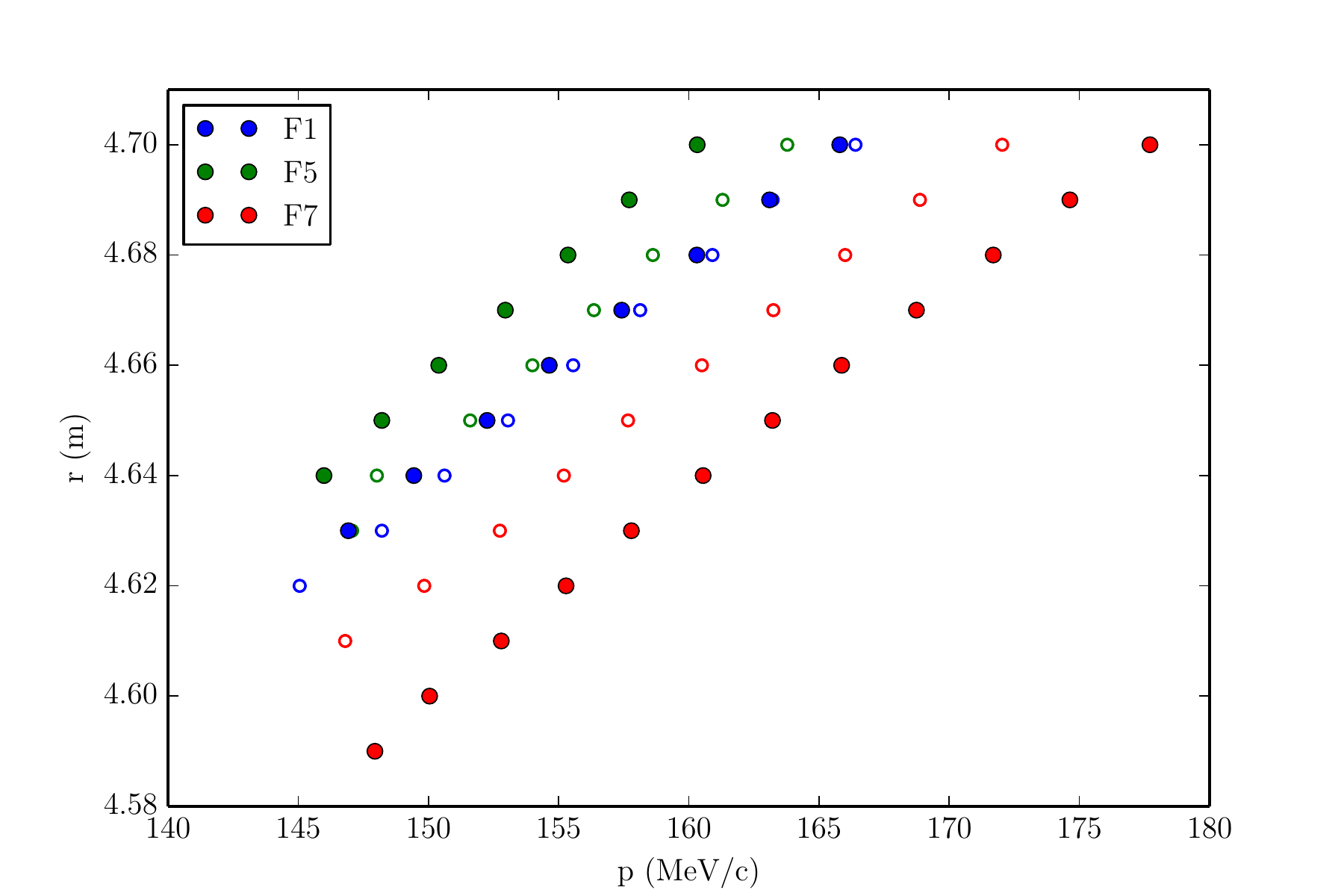}
\caption{Radial probe measurements at the three probe locations with the corrector current at 400\,A (filled circles) and at 700\,A (open circles).}
\label{fig:cod_compare}
\end{center}
\end{figure}

\begin{figure}[h!]
\begin{center}
\includegraphics[width=0.7\linewidth]{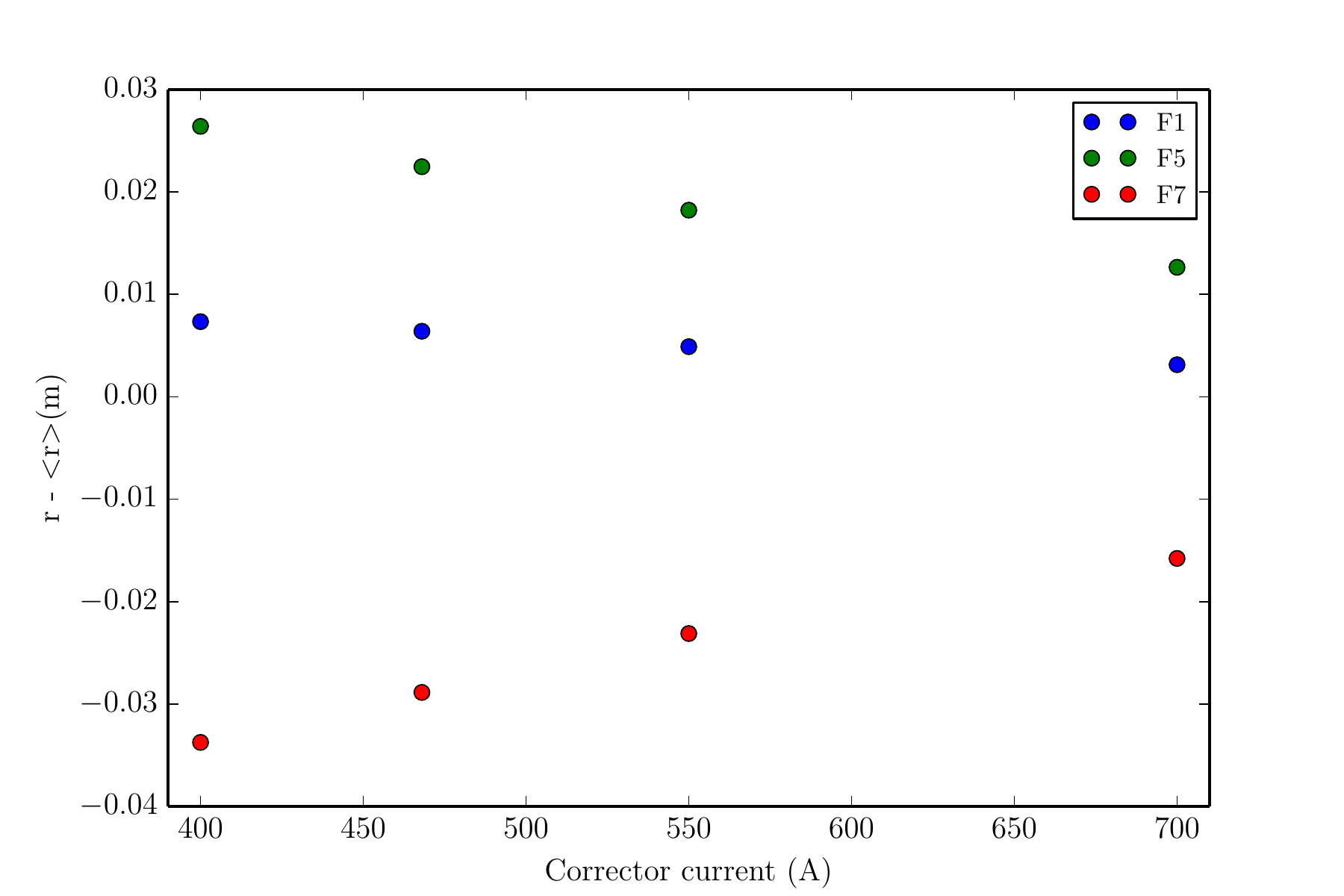}
\caption{Closed orbit distortion at the three probe locations as a function of corrector current. The COD is calculated by subtracting the mean radial beam position from that measured at each probe. The data points show at each probe the mean COD over the momentum range.}

\label{cod_measured}
\end{center}
\end{figure}

In order to operate the accelerator with optimal beam parameters, the closed orbit distortion should be minimized. One possible way to correct it is to compensate the distorted field before and after the rf cavity locally. In November 2013 corrector poles were installed to cover the full radial extent of the beam orbit excursion. Two dipole corrector poles were mounted directly before and after the rf cavity. Since the gap height of the corrector is constant, the dipole field is constant and thus the kick angle due to the correction is inversely proportional to the beam momentum.

The COD was measured in the same way as for the momentum compaction measurement in Section~\ref{sec:3.1}, using an intercepting radial probe with an accelerated beam and observing the timing when total beam loss occurs. Taking the excitation current of the corrector as a variable, the orbit position at the three probe locations was measured as a function of momentum. Radial probe measurements for the case of two corrector current settings are shown in Fig.~\ref{fig:cod_compare}. We also see in Fig.~\ref{cod_measured}, as the excitation current increases, the three curves become closer implying the asymmetry in COD is partially corrected. The results suggest that an increased corrector coil current would further reduce the COD.

\subsection{Horizontal beam size}

A measurement of the beam size is made by calculating the fall-off duration of the bunch charge monitor signal. This is similar to a method commonly used in synchrotrons~\cite{schoenauer} except without the need for a pulsed orbit distortion, as the FFAG orbit moves during acceleration. Since the signal fall-off takes place on a much slower time scale than betatron oscillations, the measurement of beam size is averaged over betatron phase. 

\begin{figure}[htb]
   \centering
   \includegraphics[width=0.7\linewidth]{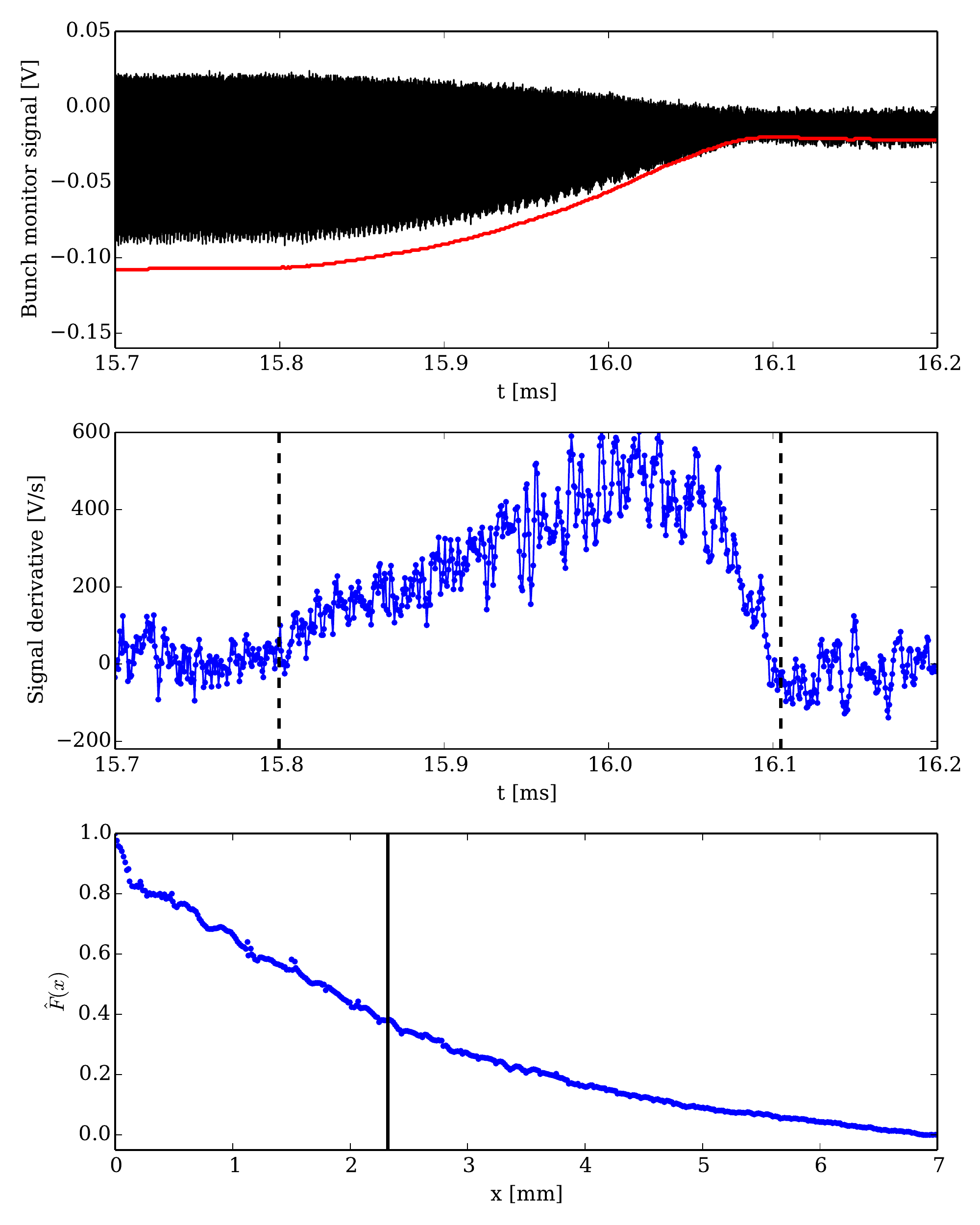}
   \caption{The top panel shows the fall-off of the bunch charge monitor signal as the beam approaches and is lost on the probe (raw data in black, moving window averaged signal amplitude in red). The centre panel shows the 
derivative of the moving window average signal amplitude. The vertical dashed lines show the duration of the signal fall-off. The bottom panel shows the distribution, normalised to maximum value, projected on to the horizontal beam axis along with the rms (solid vertical 
line).}
\label{fig:falloff}
\end{figure}

The derivative of the bunch charge monitor signal is proportional to the instantaneous number of particles scraped. The time duration within which the derivative is finite spans the time at which the bunch begins to be lost to the time when the bunch centroid reaches the probe. The derivative of the bunch charge monitor signal is shown in Fig.~\ref{fig:falloff}, a zoomed view of Fig.~\ref{fig:timetoloss} with the probe at 5.17\,m from the reference point.

Defining a density function $f$ that depends only on phase space amplitude $A$, the fall off of the bunch monitor signal $I(t)$ is given by

\begin{equation}
I(t) = 2\pi \int_0^{A_{\mathrm{max}}} f(A)AdA - 2\pi \int_{A(t)}^{A_{\mathrm{max}}} f(A)AdA = 2\pi \int_0^{A(t)} f(A)AdA,
\end{equation}

\noindent where $t=0$ is defined where $A(t)=0$ and $I(t)$ becomes zero, $A_{\mathrm{max}}$ is the amplitude limit defined by the scraper and

\begin{equation}
 X =x, X'= \beta_x x' + \alpha_x x, A^2 = X^2 + X'^2.
\end{equation}

\noindent The time coordinate can be mapped to phase amplitude coordinates via 

\begin{equation}
A(t) = \int_{p(t=0)}^{p(t)} D_{p,i}(p(t)) \frac{dp}{p(t)},
\end{equation}

\noindent where the dispersion $D_{p,i}$ and momentum function $p(t)$ obtained from the momentum compaction measurements are used. The derivative of the bunch monitor signal with respect to phase amplitude is then given by

\begin{equation}
\frac{d I(A)}{dA} = 2\pi f(A)A.
\end{equation}

Knowing the product $f(A)$ and $A$ from the measured $\frac{d I(A)}{dA}$, the Abel transformation may be used to calculate $F(x)$, i.e the distribution projected on to the horizontal axis,
\begin{equation}
F(x) = 2 \int_x^{A_{max}} \frac{f(A) A dA}{\sqrt{A^2 - x^2}}.
\end{equation}

\noindent An example of the $F(x)$ distribution obtained by numerical integration is shown in Fig.~\ref{fig:falloff}.

\begin{figure}[htb]
   \centering
   \includegraphics[width=0.7\linewidth]{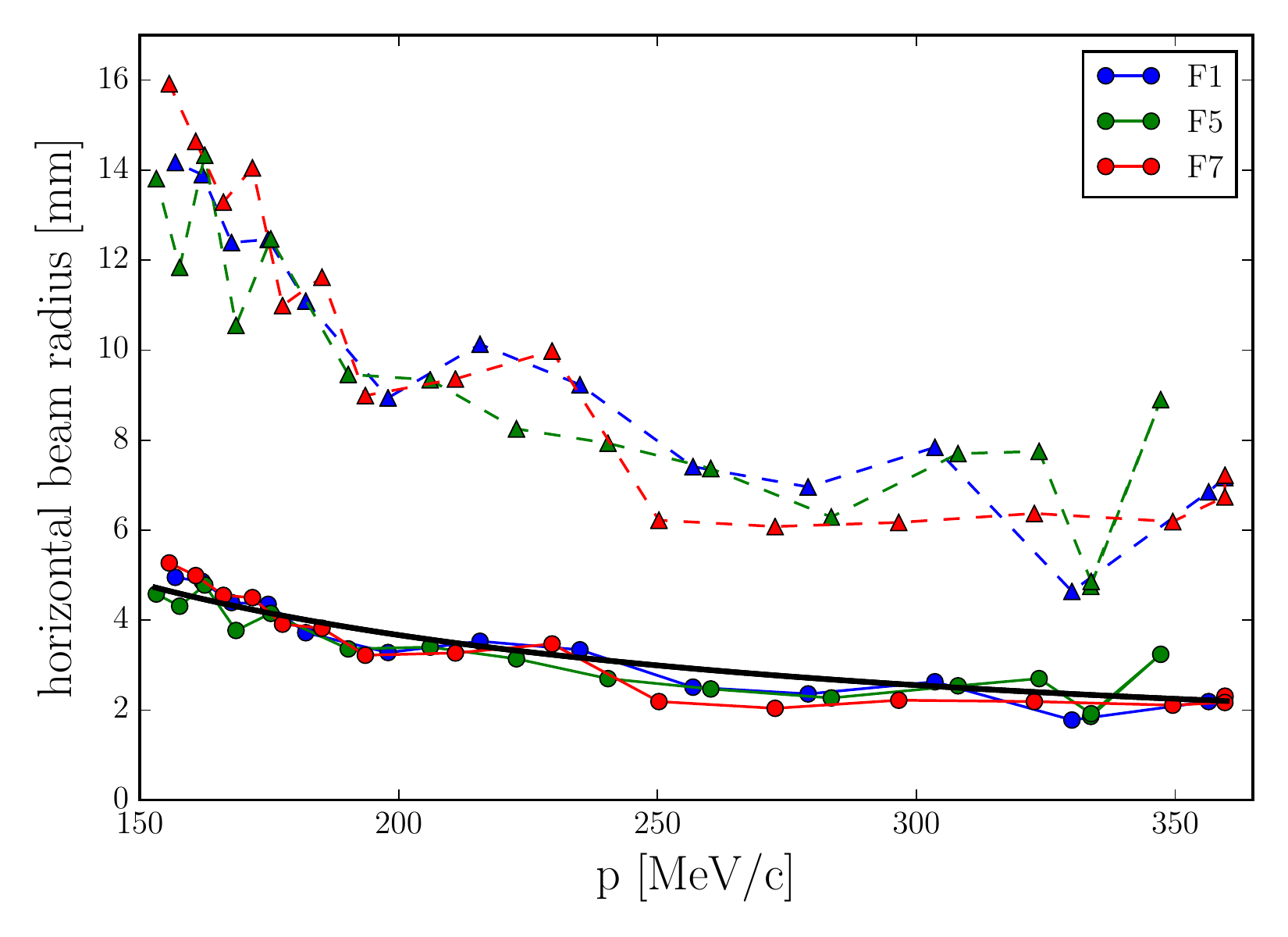}
   \caption{Horizontal beam radius found by the algorithm described in the text. Both the rms (circles, solid lines) and full extent (triangles, dashed lines) of the distribution is shown. The black line shows
 the component of the beam size arising from dispersion, assuming a momentum spread corresponding to the rf bucket height.}
 \label{fig:beamsize_results}
\end{figure}

In Fig.~\ref{fig:beamsize_results} the rms and full beam radius are shown as a function of momentum. In order to place an upper limit on the dispersive component of the beam size, a momentum spread $\delta$ corresponding to the acceptance of the rf bucket is assumed. In that case

\begin{equation}
\delta =\sqrt{\frac{2qV}{\pi \beta^2 E h |\eta|}} Y(\phi_s),
\label{eqn:bh}
\end{equation}
where $V$ is the RF voltage, $h$ is the harmonic, $E$ is the total energy of the bunch, q is the unit charge, $\beta$ is the velocity normalised by the speed of light and  $Y(\phi_s)$ is the bucket height factor~\cite{SYLee}. As can be seen in Fig.~\ref{fig:beamsize_results}, it is found that the dispersive component is of the same magnitude but smaller than the total measured beam size.

\section{Matching at injection }
\label{Section4}
In any accelerator including FFAGs, making sure the orbit and optics are properly matched at injection will minimize coherent oscillations and avoid the dilution of beam emittance. In some machines the beam is injected with a deliberate mismatch to enlarge the beam emittance in a controlled manner. This may prove necessary for high intensity operation, however the first step is to establish a method to match the beam. After this an offset may be introduced if required.

The procedure of orbit and optics matching in FFAGs is essentially the same as that in a synchrotron. However, it is important to remember that there is no ideal orbit such as the one in a synchrotron, where the orbit goes through the quadrupole centre. In the horizontal direction, the closed orbit is a function of magnet parameters even at fixed momentum without COD. It should be noted that in this FFAG we also lack the control over `beam centering' or local correction of COD, which would usually be achieved with harmonic coils in cyclotrons~\cite{Fukuda1998}.

The first orbit suitable for acceleration corresponds to the 11\,MeV injection energy (see Fig.~\ref{injec}). The main constraint of the design is that the beam has to match the equilibrium orbit at a particular reference azimuth that in our case corresponds to the foil position.

At present there is no diagnostic available to measure the optics matching. An optics model is used, based on detailed information of the lattice magnets both in the injection line and in the FFAG ring. The beam shape and emittance from the injector linac have previously been measured at one point in the injection line, which we are able to take as an initial condition for simulation work. In the following, we describe an experimental method to match the orbit in both the horizontal and vertical direction and to measure and match the horizontal dispersion at the point of the stripping foil for charge exchange injection.

\subsection{Horizontal orbit}

\begin{figure}[h!]
\begin{center}
\includegraphics[width=0.5\linewidth]{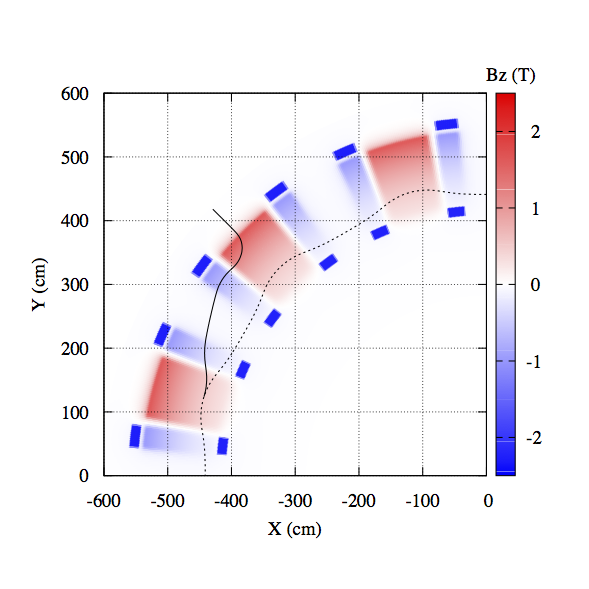}
\caption{Injection trajectory for the H- ions up to the matching point which coincides with the foil position. The first closed orbit is shown as a dotted line and the median plane field strength is indicated in coloured shading.}
\label{injec}
\end{center}
\end{figure}

The simulated injection trajectory in the main ring is shown in Fig.~\ref{injec}. The beam enters the ring after injection into the centre of one triplet and follows a complicated trajectory through the magnetic field before reaching the stripping foil. After stripping, the beam should then be on the closed orbit if the horizontal orbit is matched correctly. 

\begin{figure*}[h!]
\centerline{\subfloat[]{\includegraphics[width=0.5\textwidth]{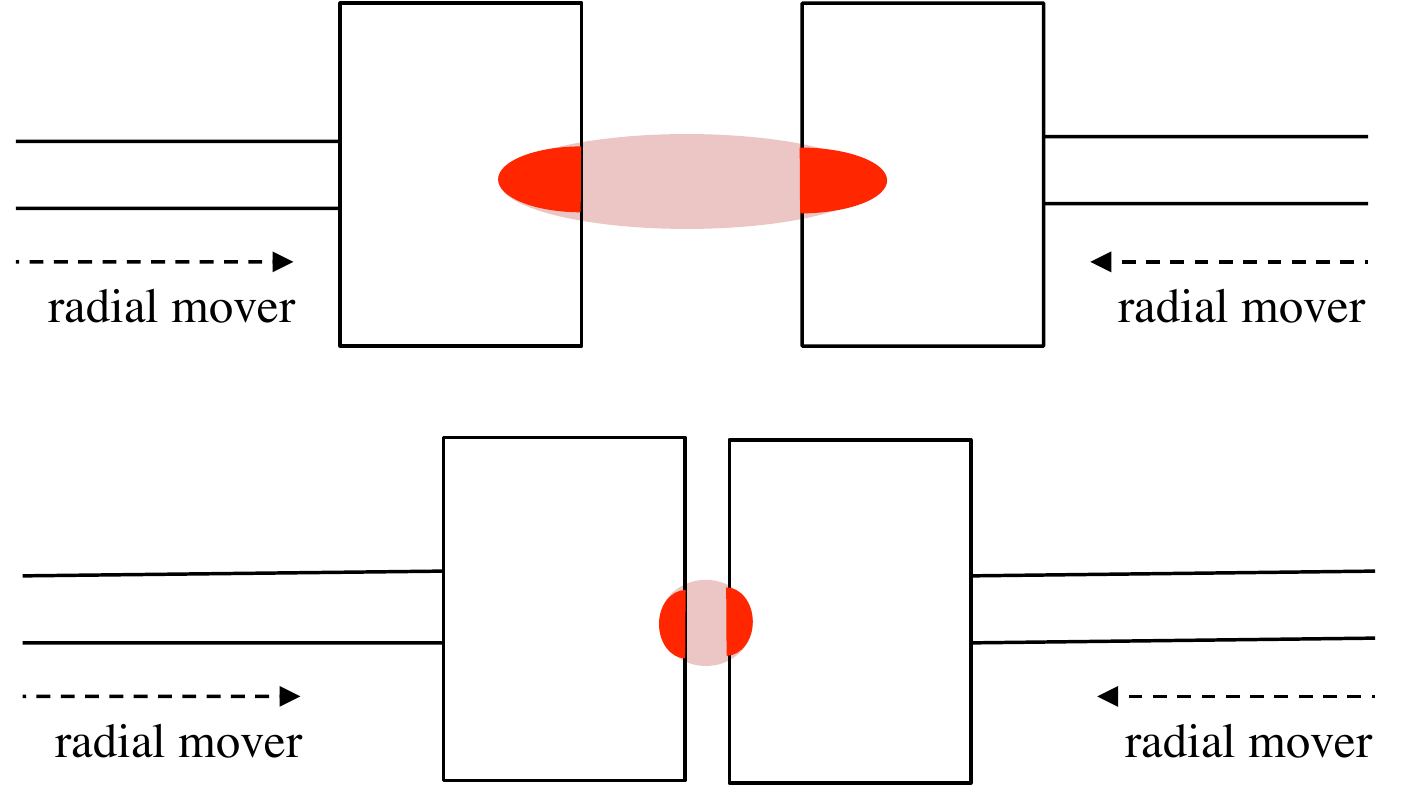}
\label{fluorescent-match}}
\hfil
\subfloat[]{\includegraphics[width=0.4\textwidth]{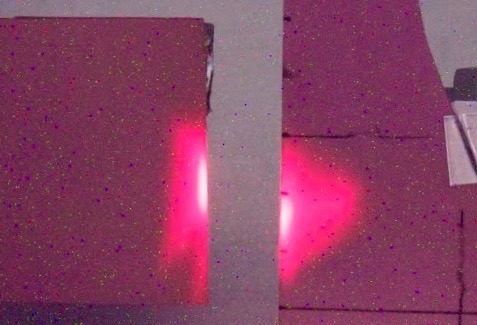}
\label{fluoro-screen}}}
\caption{(a) Layout of fluorescent screens used for orbit position matching in the horizontal plane. (b) Example of the beam on the fluorescent screens.}
\label{fig:horiz-match}
\end{figure*}

There are a number of variables which can be used to optimize the horizontal orbit matching. The present method of ensuring minimal coherent oscillations around the closed orbit is two-fold. First, a coarse adjustment is made of the foil position and injection line steering magnets to maximize the number of turns in the ring as viewed on the double plate bunch charge monitor. Next, two radially movable fluorescent screens are brought into the region of the beam, one on the inside and one on the outside. With these screens we can observe the beam size as it is scraped by the screens on successive turns. The steerers are adjusted further while bringing the fluorescent screens closer together incrementally, to achieve minimal amplitude oscillations. This procedure is performed without rf acceleration. A schematic of this setup and an example of the beam on the fluorescent screens is shown in Fig.~\ref{fig:horiz-match}.  Note that orbit matching may change as a function of the correction strength, as the closed orbit position at injection will likely change.

Ideally, the radially movable beam position monitor would be used to measure the amplitude and minimize horizontal coherent oscillations in a non-destructive manner, however it was not installed during this measurement. In fact, if two of these devices were located in a single straight section it would allow both the position and gradient to be measured at injection. Nevertheless, the present method allows a reasonable level of confidence that the horizontal orbit is well matched.

\subsection{Vertical orbit}

In the vertical direction, the target orbit is zero in both position and gradient. To minimize any small mismatch small adjustments are required, which can be controlled by the steering magnets in the injection line. There are three steering magnets (named S4V, S4PV and S5V) and the phase advance between S4PV and S5V is almost 90 degrees, which allows us to use them as independent controls. The vertical matching was also performed without rf acceleration.

Coherent dipole oscillations were observed using the beam position monitor half way downstream around the ring in S12. In order to find the optimum injection in the vertical direction, two vertical steering magnets (S4PV and S5V) were scanned in the injection line with the aim of minimizing the coherent oscillations observed in the ring.

\begin{figure*}[!h]
\centerline{\subfloat[]{\includegraphics[width=0.5\textwidth]{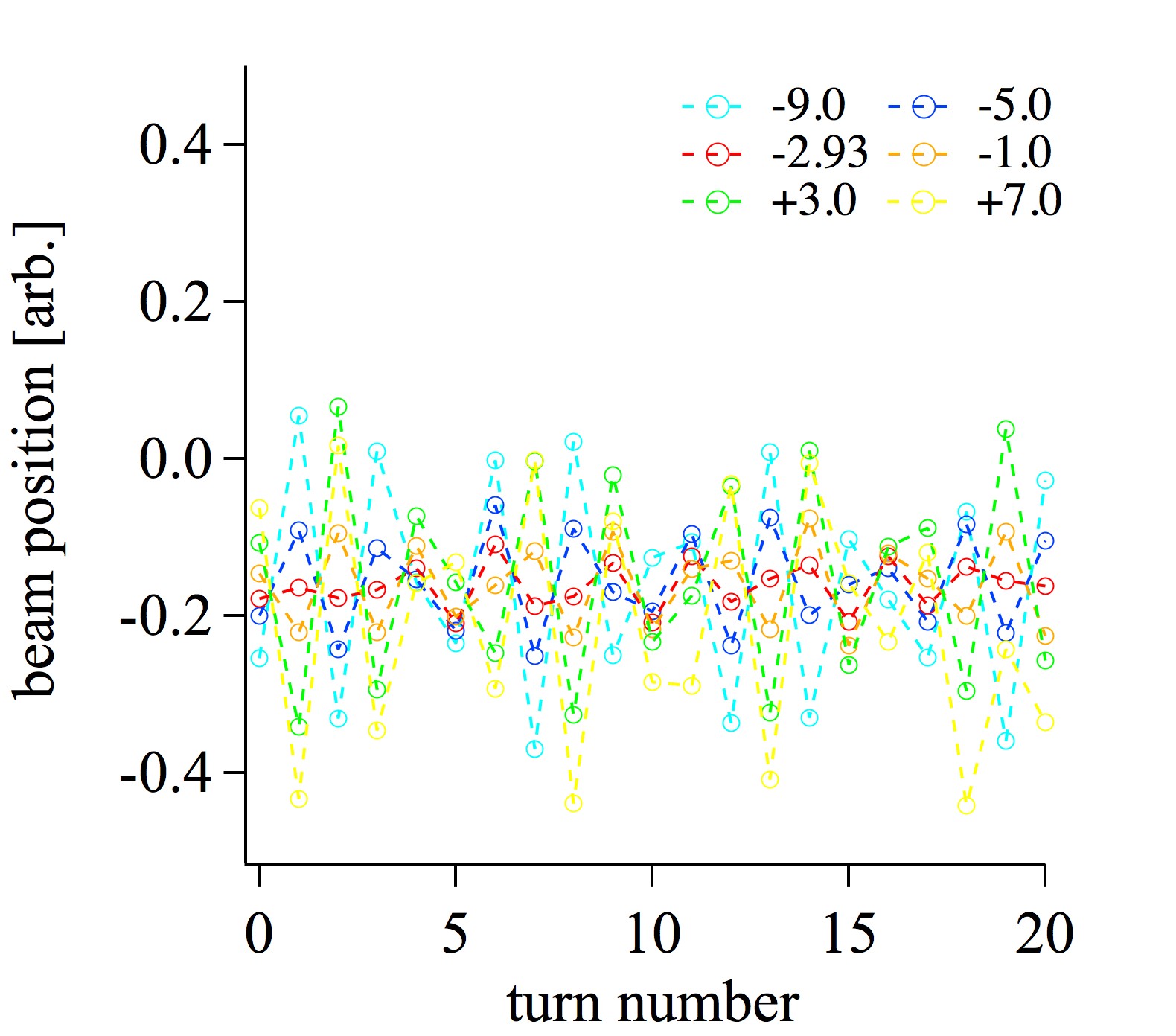}%
\label{D1012_ST5V}}
\hfil
\subfloat[]{\includegraphics[width=0.5\textwidth]{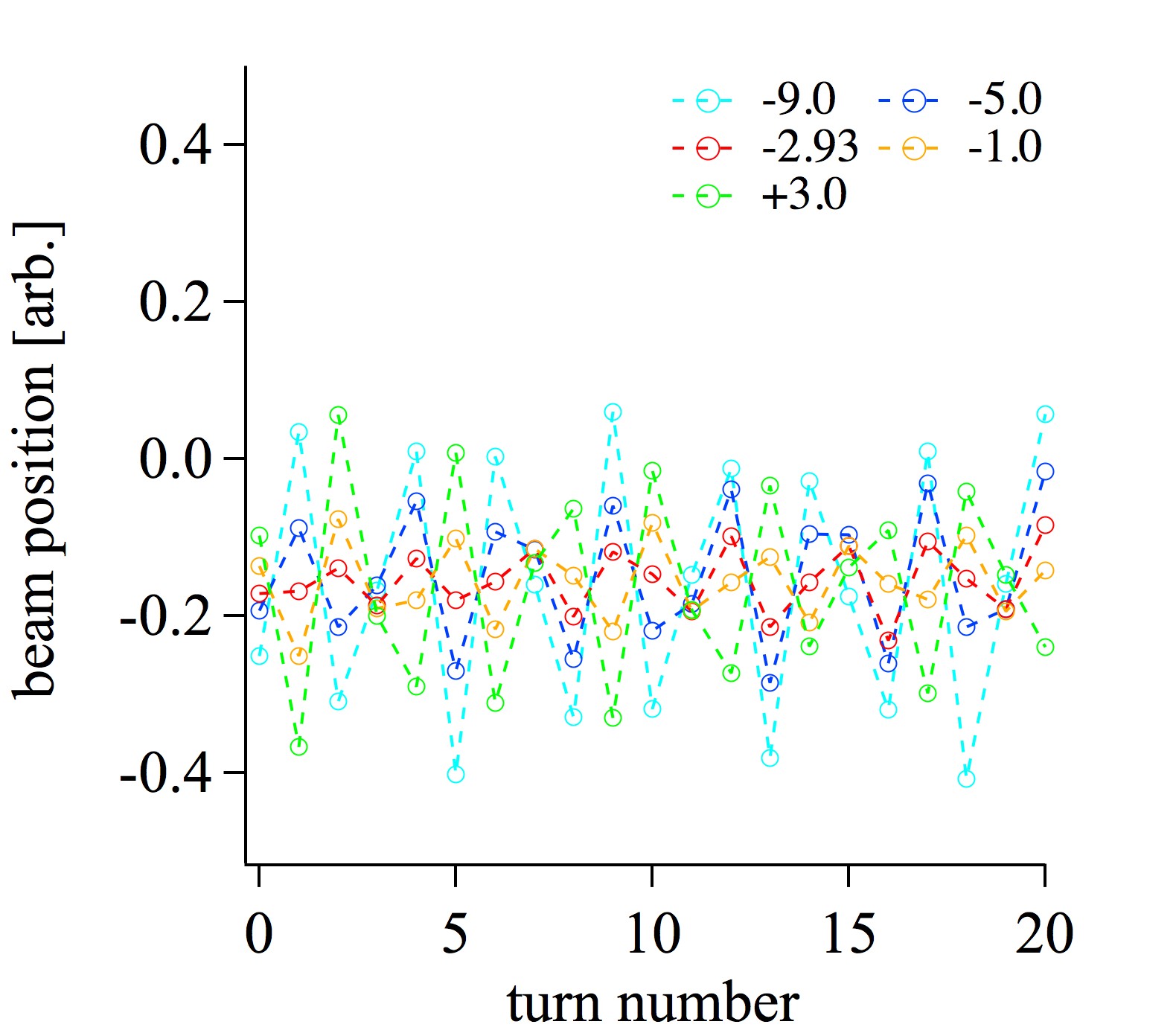}%
\label{D950_ST5V}}}
\caption{Coherent oscillations of the first 20 turns observed for varying steering magnet current of S5V for two working points, (a) D=1012\,A and (b) D=950\,A. The unit of the legend is [A].}
\label{fig:vert_matching}
\end{figure*}

We first scanned the main D magnet current from 850\,A-1200\,A with a view to measuring the vertical tune variation.\footnote{In order to see the coherent oscillations within the small number of turns, the fractional tune must be set away from an integer.} We next fixed the D magnet at the nominal current of 1012\,A and scanned one vertical steering magnet S5V. As can be seen in Fig.~\ref{fig:vert_matching}\,(a), the minimum amplitude of vertical coherent oscillations corresponds to the empirically optimized steering current of -2.93\,A.

The measurement was repeated with the main D magnet current set to 950A, thus with a different vertical tune. Figure~\ref{fig:vert_matching}\,(b) confirms that the same steering magnet current gives the minimum coherent oscillation amplitude in this case also.

\begin{figure}[h!]
\begin{center}
\includegraphics[width=0.5\textwidth]{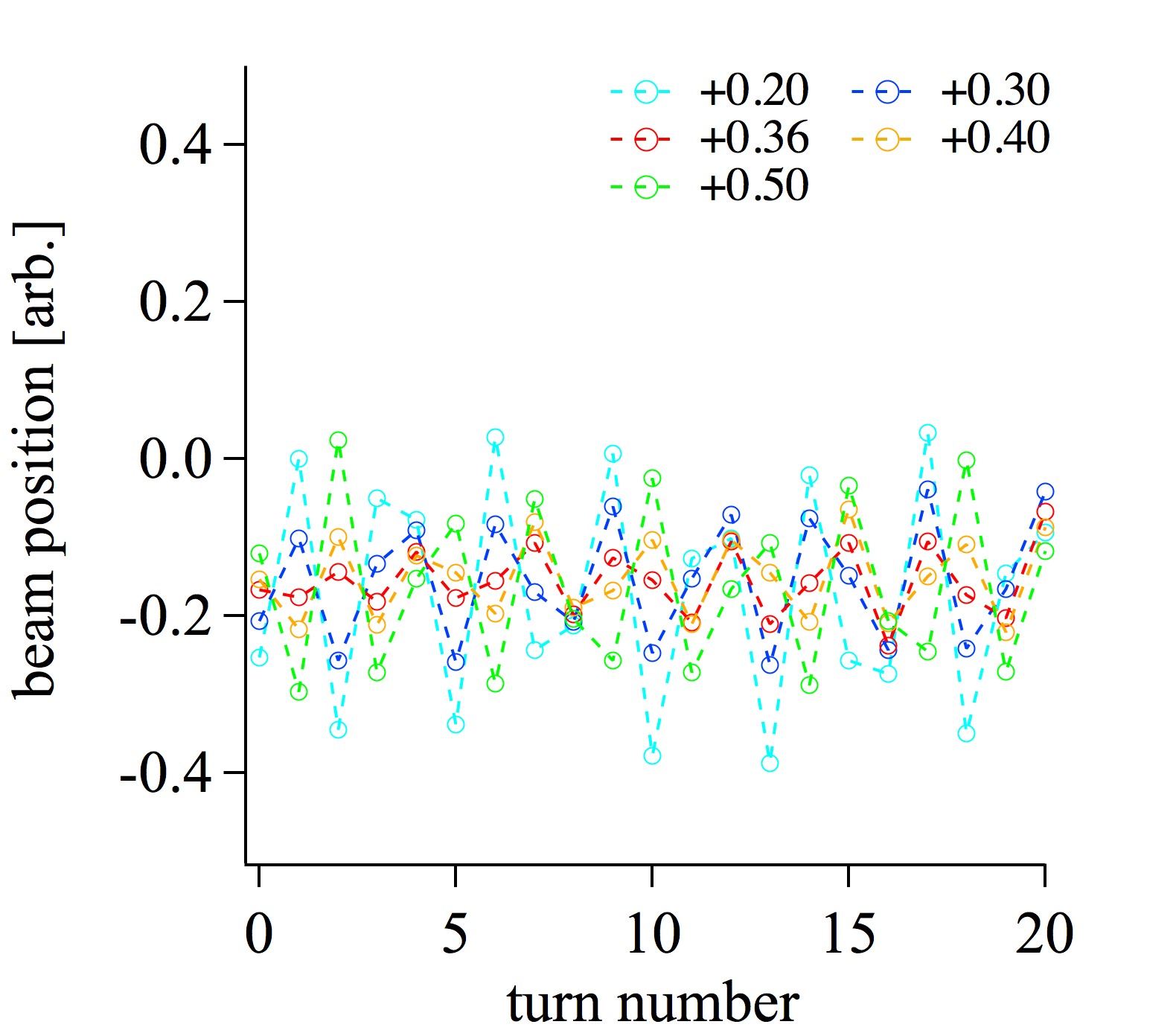}
\caption{Coherent oscillations of the first 20 turns observed for varying steering magnet currents of S4PV. Unit of the legend is [A].}
\label{D950_ST4pV}
\end{center}
\end{figure}

Steering magnet S4PV is an orthogonal control to S5V. The results of optimising the strength of this magnet are shown in Fig.\ref{D950_ST4pV}. The current setting of +0.36\,A produces the minimum oscillation amplitude in this case. In the process of scanning steering magnets, we noticed that the oscillation centre is not zero, but slightly negative. This indicates that there may be COD in vertical direction. Since there is no other vertical BPM in the ring, it is not possible to confirm whether this is due to COD, however other indications such as beam loss hint that this might be the case. It is hoped this can be confirmed and if necessary corrected when suitable hardware is available.

\subsection{Dispersion function}

\begin{figure}[h!]
\begin{center}
\includegraphics[width=0.8\textwidth]{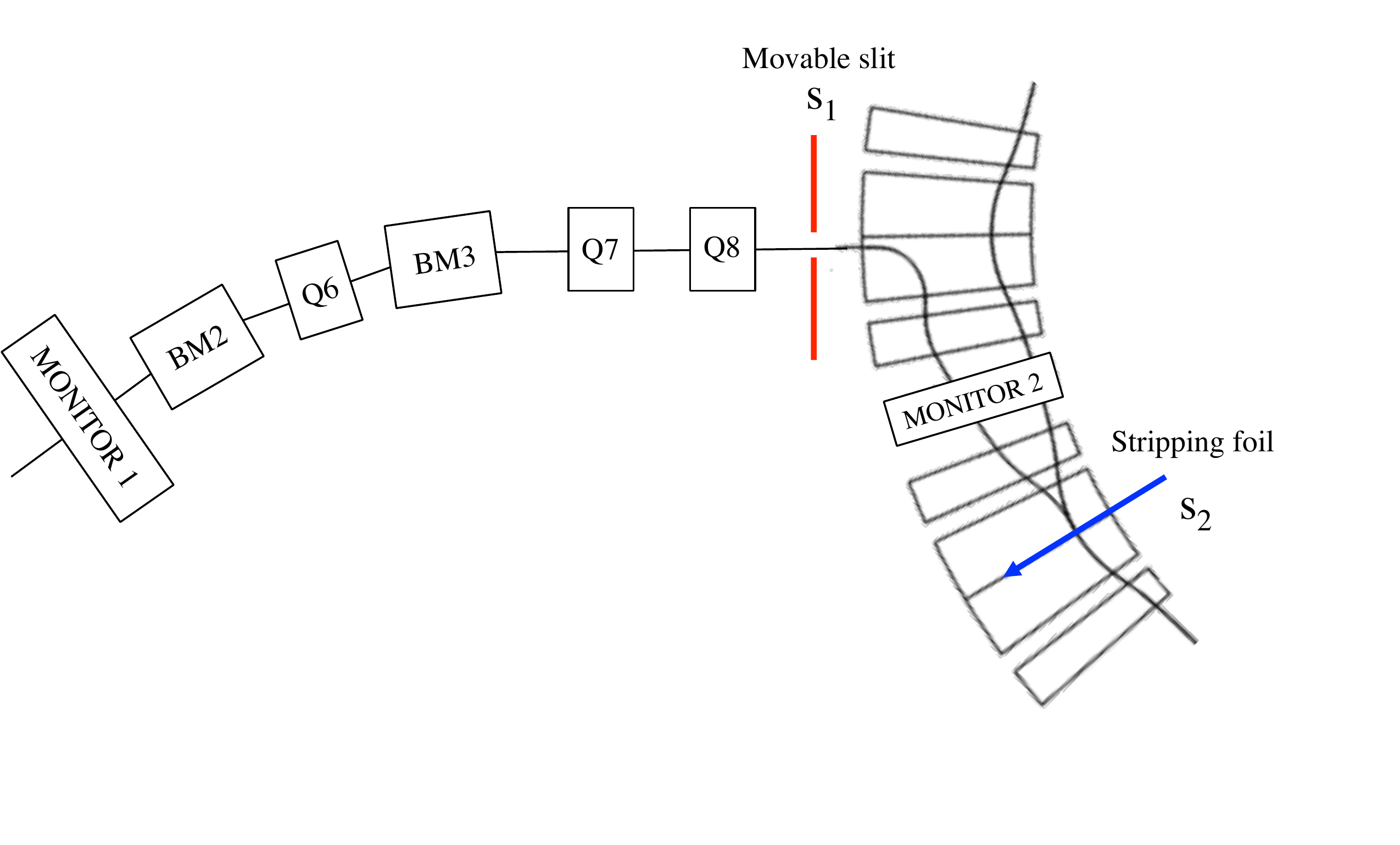}
\caption{Injection line layout of magnets used for dispersion function measurement at the movable slit position $s_{1}$, where the beam from the linac transfer line is injected into the ring using the stripping foil at $s_{2}$. Two cells of the FFAG ring are shown on the right for illustration. Note that five additional quadrupoles exist upstream of Monitor 1.}
\label{fig:injectsetup}
\end{center}
\end{figure}

Particles with a relative momentum deviation $\frac{\Delta p }{p_0} \ne 0$ will follow a different path from the orbit of an on-momentum particle, as they experience a different amount of bending. This dispersion is created in the first instance in the injection transfer line by dipole magnets, as seen schematically in Fig.~\ref{fig:injectsetup}. Naturally the total beam size in the ring will consist of a component due to betatron oscillations and a component due to dispersion, motivating the measurement and understanding of dispersion matching in this machine.

The basic dispersion function measurement method is to determine how the beam moves transversely as a function of momentum. For the ring, we have already deduced the periodic dispersion function from the orbit measurement with momentum that was carried out in Section~\ref{Section3}, which is around $0.59$\,m at injection. The method used is similar to the rf frequency shift method of synchrotrons~\cite{minty03}. In this section we aim to develop a method to match the dispersion function of the injection line to the periodic dispersion function in the FFAG ring.

The layout of the injection line is shown in Fig.~\ref{fig:injectsetup}. There are two measurement points in the injection line which will be discussed in this section, the first is upstream from the stripping foil at the position of a movable slit following the final quadrupole in the injection line, which we call $s_{1}$. The second measurement point is at the position of the stripping foil itself, which we refer to as $s_{2}$.

To measure the dispersion function at the location of the stripping foil $s_{2}$, we can use two methods. The first is a direct measurement method varying the momentum of the beam coming from the injector linac. Initial results using this method did not agree with the measured periodic dispersion function in the ring, suggesting that at present the dispersion function is not matched. Later in this section a campaign is described which measures the dispersion vector at a number of different sections using the `equivalent momentum' method to find the source of this discrepancy. Control and matching of the dispersion function at the stripping foil are then discussed.

\subsubsection{Direct dispersion function measurement at foil}
%  Method and result are from Ishi-san's note 2015.06.29
We vary the tank field in the second drift tube tank of the injector linac (DTL2) in order to change the beam momentum to measure the dispersion function at $s_{2}$, $D(s_{2})$ at the foil. Five different tank field levels are used, called case 1 to case 5. The time-of-flight difference in a 16.5\,m straight section of the beamline from the end of the linac to Monitor 1 (M1) is measured for each setting. This is used to determine the momentum deviation as $\Delta p/p=-\gamma^{2}(\Delta T/T)$, shown in Table~\ref{table:momdev}. Case 4 represents the nominal values with an energy of around 11\,MeV and a time-of-flight of 362\,ns.

\begin{table}[h!]
\caption{Momentum deviation from variation of DTL2 tank field.}%%%Table caption goes here
\label{table:momdev}
\centering
\begin{tabular}{|c||c|c|c|c|c|}%%%The number of columns has to be defined here
\hline
& case 1 & case 2 & case 3 & case 4 & case 5 \\
\hline
$\Delta T$ [ns] & 12.7 & 8.3 & 4.0 & 0.0 & -2.7\\%%%% Table body
\hline
$\Delta p/p$ [\%] & -2.6 & -1.7 & -0.8 & 0.0 & 0.5\\
\hline
\end{tabular}
\end{table}%%%End of the table

\begin{figure*}[!h]
\centerline{\subfloat[]{\includegraphics[width=0.5\linewidth]{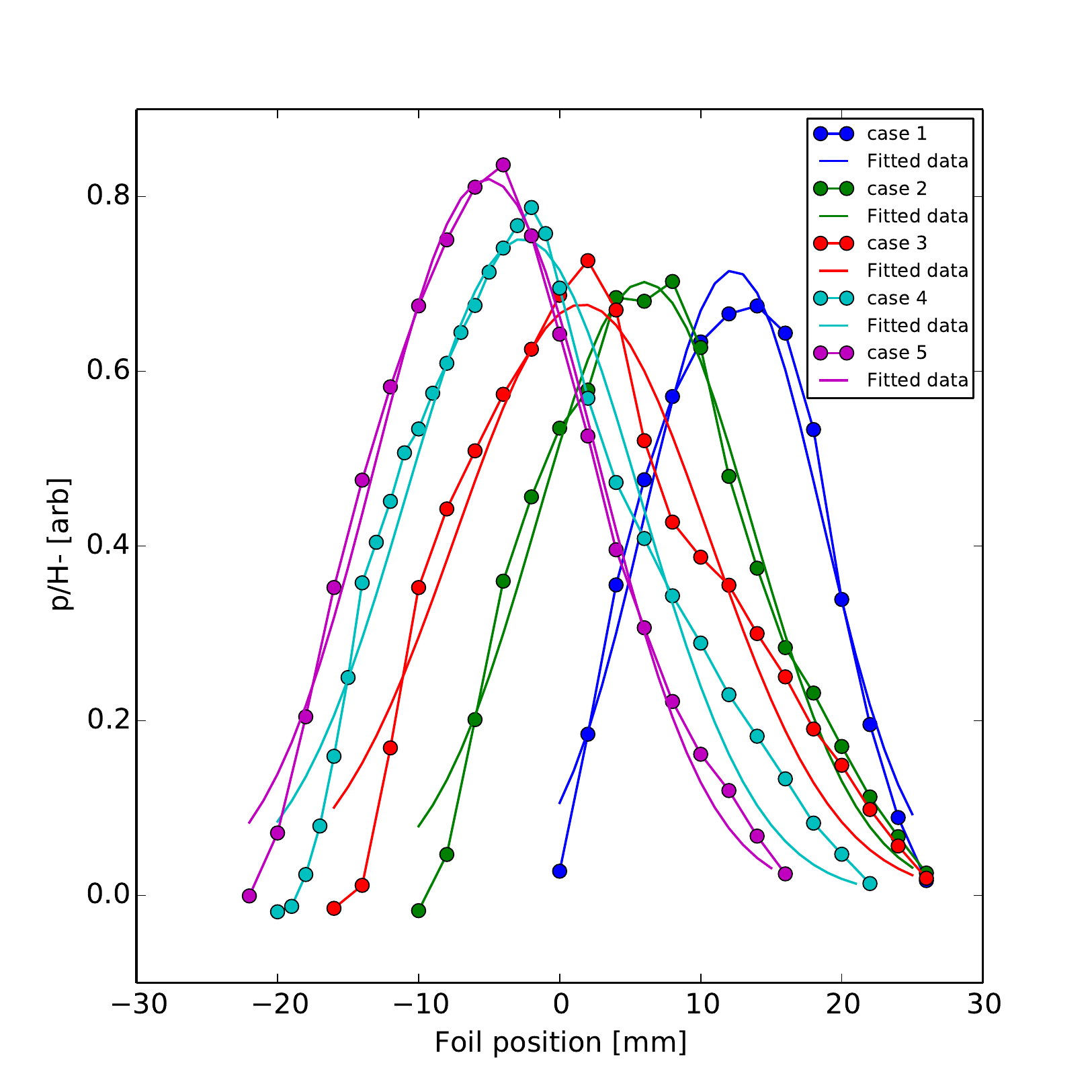}%
\label{fig:disp_direct_fit1}}
\hfil
\subfloat[]{\includegraphics[width=0.5\linewidth]{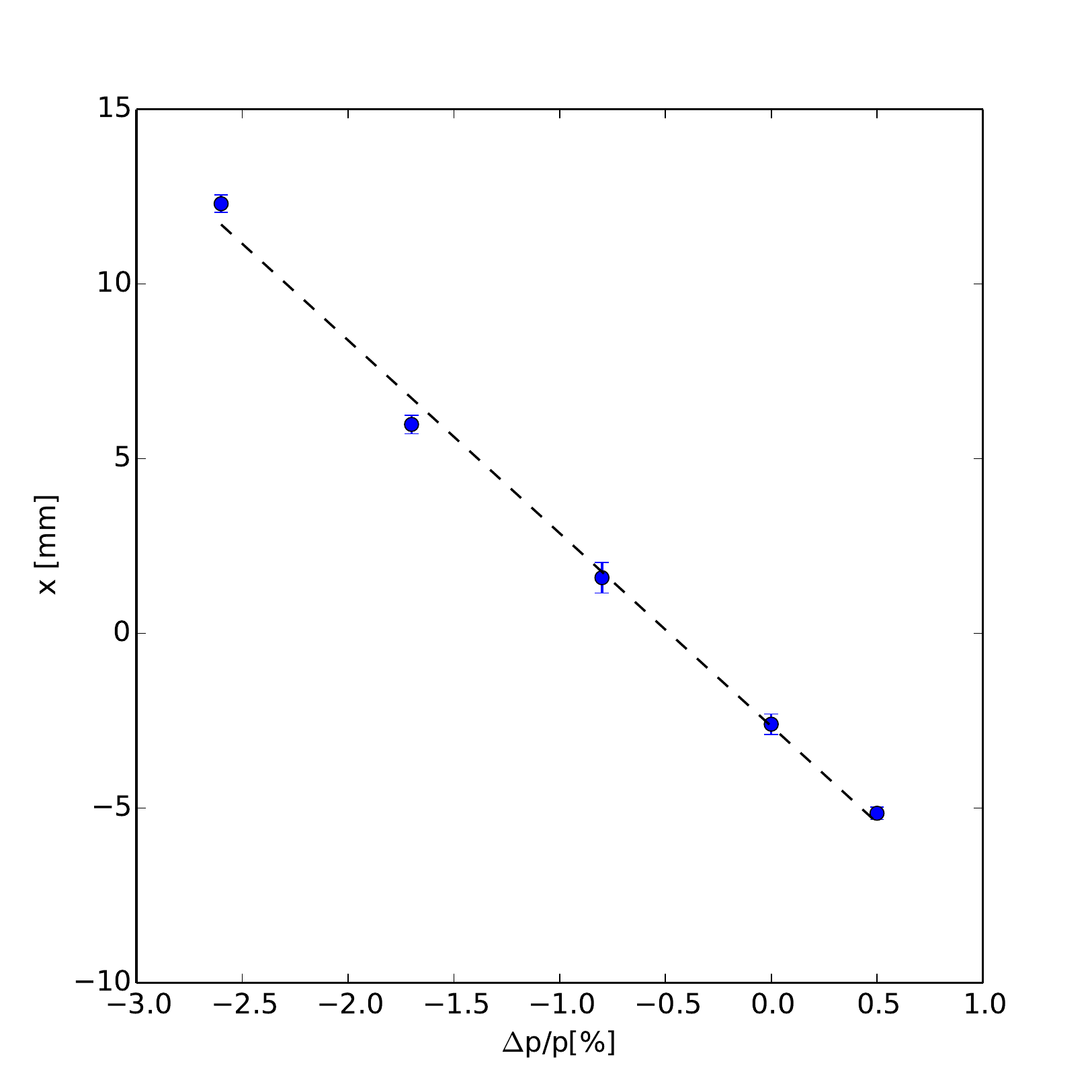}%
\label{fig:disp_direct_fit2}}}
\caption{(a) Measured bunch charge monitor peak ratios and (b) the resulting dispersion function measurement at $s_{2}$, the location of the stripping foil. }
\label{fig:dispersion_direct}
\end{figure*}

We determine the relationship between foil position and conversion ratio from H$^{-}$ to protons using the bunch charge monitor in S7 indicated as Monitor 2 (M2) in Fig.~\ref{fig:injectsetup}. In this instance we compare the incoming H$^{-}$ bunch charge to the bunch charge of protons after one turn at the same monitor. When the incoming beam is centred on the foil, the ratio of protons converted from H$^{-}$ is at a maximum. On the other hand, when the foil is located at the edge of the incoming beam, the proton ratio is lower. The relationship between conversion ratio from H$^{-}$ to protons and the foil position gives the beam profile at the foil~\footnote{Note that the measured profile using the foil is not the true beam profile, but rather the profile elongated by the foil width. The profile itself is not used in this measurement, only the central position.}. We use a Gaussian profile to fit the measured data to determine the central position for each momentum setting, as shown in Fig.~\ref{fig:disp_direct_fit1}. Using a linear regression, the measured value of the dispersion function at the foil is $D(s_{2})=-0.55\pm0.03$\,m shown in Fig.~\ref{fig:disp_direct_fit2}. 

This result suggests the dispersion function is not matched at the position of the stripping foil. To find the reason for the discrepancy, we need to obtain information about the dispersion earlier in the injection line, so that we can establish where the mismatch arises with comparison to the design values. 

\subsubsection{Dispersion vector measurement}
\label{injline_disp}

To investigate the discrepancy of the dispersion function of the injection line at the foil position, we measured the dispersion vector locally in the injection line, not the dispersion function itself. Firstly we measured the dispersion vector from $s_1$ to $s_2$ and saw if it propagated the dispersion function upstream properly to the foil position.

We employ a method which does not require any adjustment of the injector (In fact, we want to fix the optics and the beam parameters before $s_1$ to extract the dispersion vector from $s_1$ to $s_2$). This is the `equivalent momentum' method. The method relies on the fact that changing the current in the main magnets (while keeping the ratio of all the magnetic field strengths fixed) will move the beam position in an inverse manner to the case where the beam momentum is changed. In other words decreasing the main magnet excitation currents will bend the beam less in the same way that an increase in momentum would bend the beam less. A similar method was successfully employed in the commissioning of EMMA~\cite{Barlow2010c, Machida2012a}. 

One complication in this particular machine is that the field response of the return-yoke free magnets does not necessarily follow the excitation current in a linear way. To determine the required magnet excitation current the conversion between applied current and change in magnetic field was calculated using the 3D magnet model in TOSCA. This method is described in more detail in Appendix~\ref{appendixa}.

To find the beam position at $s_{2}$ as a function of equivalent momentum, the ratio of the beam current of protons after the foil to the beam current of H$^{-}$ at M2 (a half cell before the foil) is measured as a function of foil position. As before, the relationship between conversion ratio from H$^{-}$ to protons and the foil position gives the beam profile at the foil. We use a Gaussian profile to fit the measured data to determine the central position. The process was repeated for each equivalent momentum. The result is shown in Table~\ref{dispersion_foil}.

\begin{table}[!h]
\caption{Dispersion vector measurement from $s_1$ to $s_{2}$.}%%%Table caption goes here
\label{dispersion_foil}
\centering
\begin{tabular}{|c||c|c|}%%%The number of columns has to be defined here
\hline
& $d(s_{1}-s_{2})$ [m] & error [m] \\
\hline
Nominal setup & -0.59 & $\pm 0.07$\\%%%% Table body
\hline
\end{tabular}
\end{table}%%%End of the table

Once we know the dispersion vector from $s_1$ to $s_2$, we can calculate the dispersion function at the foil with the transfer matrix from $s_1$ to $s_2$ and the dispersion function at $s_1$, as in Eq.~(\ref{eqn:dispersion}).\footnote{Here we follow the notation in p.\,116 of Ref.~\cite{SYLee}.}

\begin{equation}
\left[ \begin{array}{c} D(s_{2}) \\ D'(s_{2}) \end{array} \right] = \begin{bmatrix} M_{11} & M_{12}  \\ M_{21} & M_{22} \end{bmatrix} \left[ \begin{array}{c} D(s_{1}) \\ D'(s_{1}) \end{array} \right]
+ \left[ \begin{array}{c} d \\ d' \end{array} \right]
\label{eqn:dispersion}
\end{equation}

Taking the simulated transfer matrix and the measured value of $d(s_{1}-s_{2})$ together with the simulated values of dispersion at $s_{1}$ for the injection line setup used during the foil measurement, we obtain:

\begin{equation}
\arraycolsep=3pt
\left[ \begin{array}{c} D(s_{2}) \\ D'(s_{2}) \end{array} \right] = \begin{bmatrix} -3.36 & -0.89 \\ -2.51 & -0.96 \end{bmatrix} \left[ \begin{array}{c} -0.98 \\ 2.45  \end{array} \right] + \left[ \begin{array}{c} -0.59\pm 0.07 \\ d'(s_{1}-s_{2}) \end{array} \right]
 = \left[ \begin{array}{c} 0.54 \pm 0.07 \\ 0.098 + d'(s_{1}-s_{2}) \end{array} \right]
\label{eqn:dispersioncheck}
\end{equation}

In other words, as long as the injection line optics are the same as the simulated values, the dispersion function should be matched at the position of the stripping foil. This gives us confidence that our understanding of the optics from the entrance at $s_{1}$ to the foil at $s_{2}$ is accurate.

Secondly we measured the dispersion vector from M1 to $s_1$ with the equivalent momentum method. There is no bending magnet before M1.
To measure the dispersion vector from M1 to $s_{1}$, we scale all the injection line magnets after M1 together at the nominal current setting (0\%) and at -2\%, -1\%, +1\% and +2\%. This is more straightforward than in the main ring as the excitation of the injection line magnets is linear. At each equivalent momentum setting, we move the slit systematically across the width of the beam, and record the bunch charge monitor signals on both M1 (upstream) and M2 (which is located in the ring before the stripping foil). The ratio of the beam current at M2 over M1 as a function of the slit location gives the transverse beam profile at $s_{1}$. 

We use a Gaussian profile to fit the measured data as before to determine the central position, which is defined as the beam position. The process was repeated for each equivalent momentum. From this we can calculate the dispersion vector using a linear regression. Note that the form of Eq.~(\ref{eqn:dispersion}) applied at $s_{1}$ indicates that the measured dispersion vector is equal to the dispersion function at $s_1$ because there is no bending magnet before M1 and the dispersion function and its derivative at M1 are zero.
The measured dispersion vector compared to the design value at $s_{1}$ is shown in Table~\ref{dispersion_s1}.

\begin{table}[h!]
\caption{Dispersion function measurement from M1 to $s_{1}$.}%%%Table caption goes here
\label{dispersion_s1}
\centering
\begin{tabular}{|c||c|c|c|}%%%The number of columns has to be defined here
\hline
& $d(M1-s_{1})$ [m] & error [m] & design value \\
\hline
Nominal setup & -0.36 & $\pm 0.07$ & -0.98\\ %%%% Table body
\hline
\end{tabular}
\end{table}%%%End of the table

\begin{figure}[h!]
\begin{center}
\includegraphics[width=0.7\linewidth]{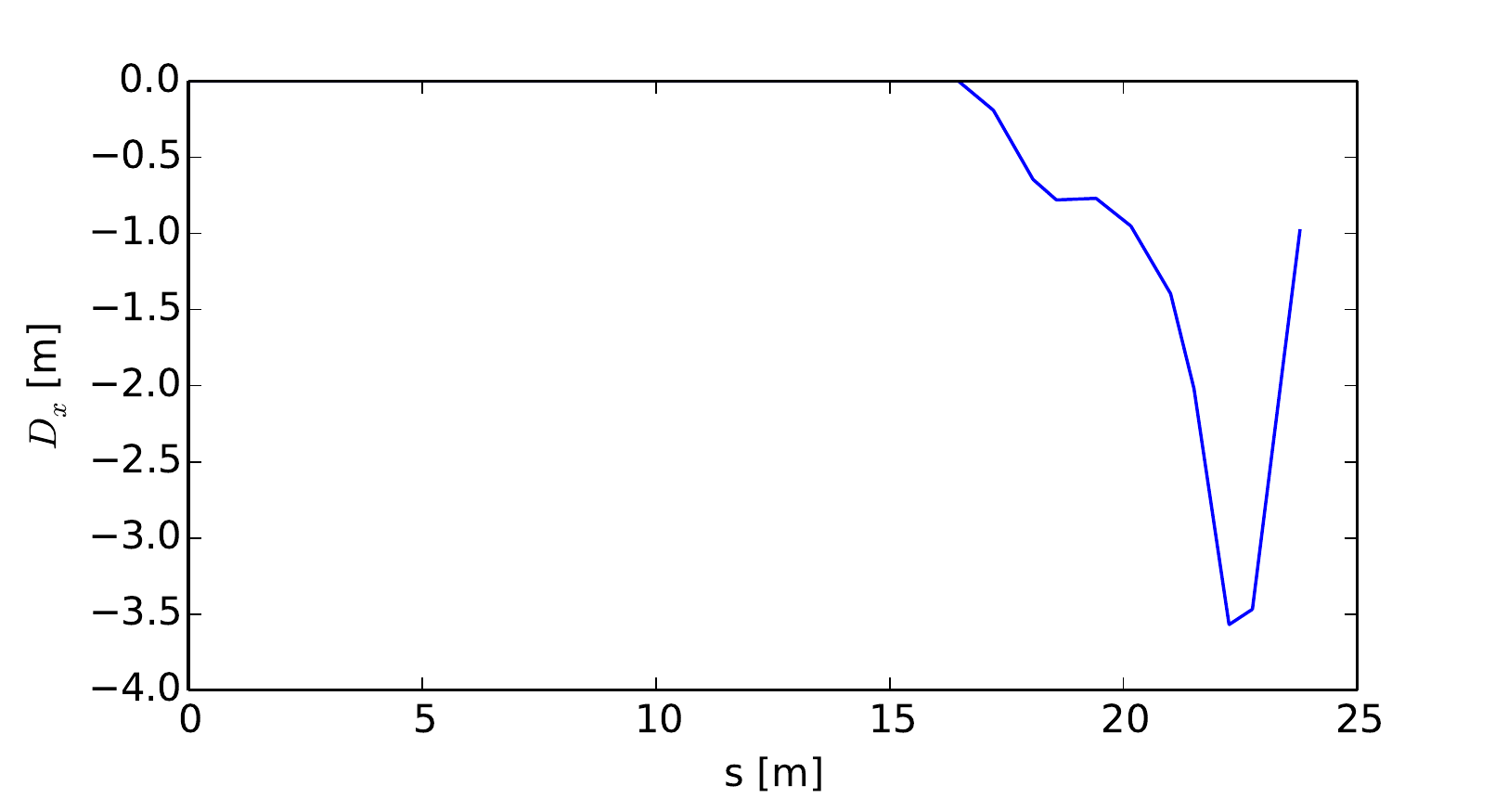}
\caption{Model of dispersion in the injection line before the measurement point $s_{1}$. Note that the gradient of the dispersion at the end of this section is large.}
\label{fig:injline_model}
\end{center}
\end{figure}

This results in the dispersion function $s_{1}$ that is different from the design value. From this we can conclude that the source of the discrepancy of the dispersion function at the stripping foil is the optics before $s_1$. This can be understood as the optics of the injection line are designed in such a way that there is a large value of the gradient of the dispersion function at $s_{1}$. Thus a small change of the quadrupole parameters can vary the optics dramatically, easily changing the dispersion function at $s_{1}$ and thus the magnitude and/or sign of the dispersion function at the stripping foil. As an example, the model of the injection line dispersion using MADX\footnote{\url{http://madx.web.cern.ch/madx/}} is shown in Fig.~\ref{fig:injline_model}.

We know that there are small variations between individual injection line quadrupoles, which were re-used from a previous experiment and currently assume a common, identical excitation curve. An opportunity to remove and measure the excitation curves of each individual quadrupole is currently being sought. In the meantime, with three injection line quadrupoles available, it will be possible to properly match the dispersion function at the location of the foil to the periodic dispersion function in the main ring\footnote{The additional five quadrupoles upstream of M1 may be used to match the Twiss parameters. Additional wire scanners would be needed to complete this full optics measurement and matching scheme.}.

\section{\label{Section5} Measurement of betatron tune}

In the ideal case a scaling FFAG has static betatron tunes throughout the whole acceleration cycle. In reality the tunes may have some small variation, primarily because it is impossible to create a magnet which produces a perfect field profile. It has been observed in operation of the ADSR-FFAG that there are a number of key loss points in the acceleration cycle. One of the aims of measuring the betatron tunes in detail is to start to identify which resonance lines these loss points might correspond to. 

The methods for measuring horizontal and vertical tunes are slightly different because of the radial movement of the beam with acceleration. There are three main regions to consider for horizontal tune measurements. For the majority of the energy range the horizontal betatron tunes are measured using the radially movable rf perturbator device and the radially movable triangle plate BPM. 

When the beam is in the injection region and in particular when it is still passing through the injection foil this method is difficult to employ. However, in the low energy region an additional frequency can be superimposed on the rf system to apply a coherent oscillation that can facilitate tune measurements in the early parts of the acceleration cycle. At higher momentum when the beam is near extraction, the radius is out of range of the radial movers and so in this region the kicker magnets are used to perturb the beam horizontally. 

The vertical betatron tune is measured using the vertical perturbator located in S3 together with the double plate bunch charge monitor. In both the horizontal and vertical case, the coherent oscillations induced by the perturbating devices are observed as sidebands in the Fourier spectrum of the bunch charge monitor signal using a real time spectrum analyser. While the hardware used has been adapted for the large aperture of the FFAG, the tune measurement method is similar to that used in other machines~\cite{minty03}.

\begin{figure}[h!]
\begin{center}
\includegraphics[width=0.7\linewidth]{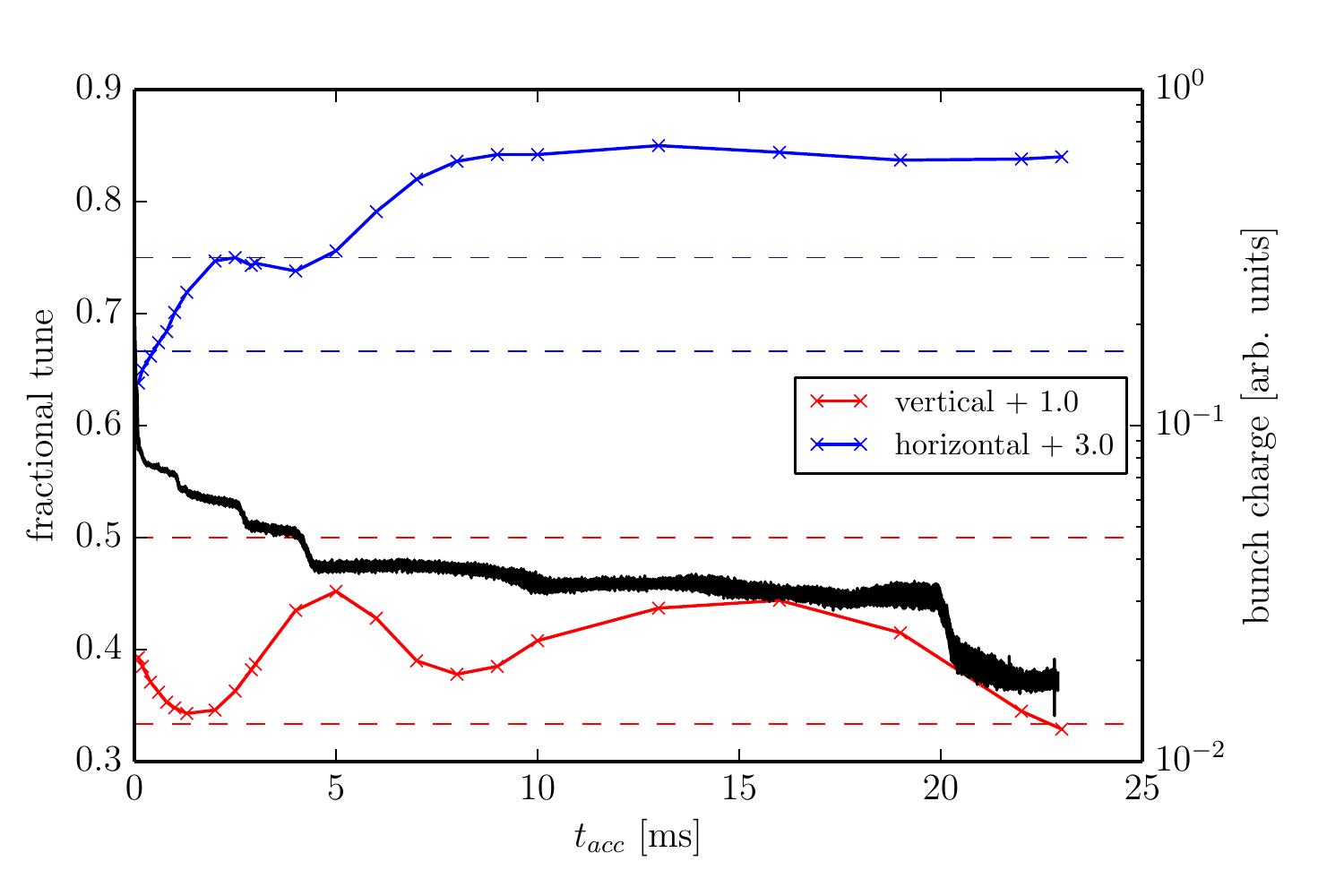}
\caption{Measured fractional betatron tunes for the working point $I_F=814$ [A] and $I_D=1012$ [A] throughout the acceleration cycle. Horizontal lines indicate nearby structure resonances. Measured beam loss is shown in black points.}
\label{Figure21}
\end{center}
\end{figure}

\begin{figure}[h!]
\begin{center}
\includegraphics[width=0.6\linewidth]{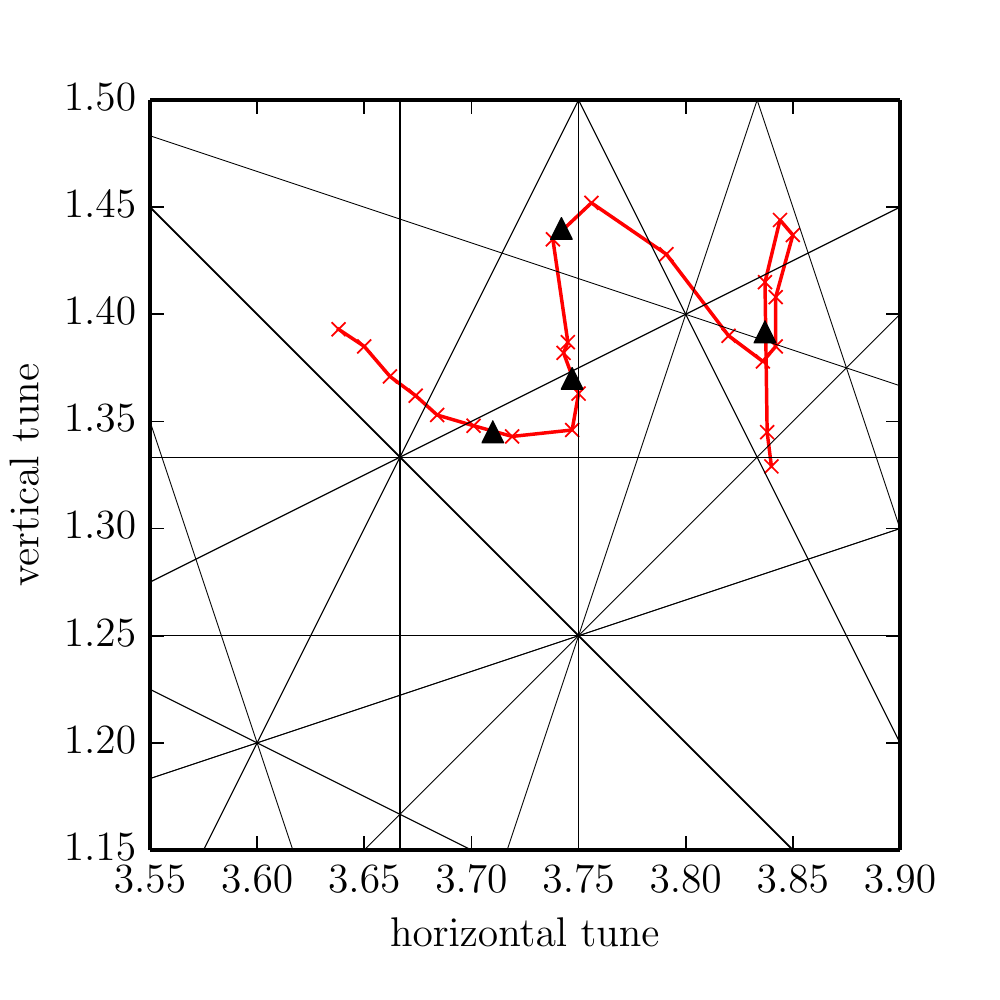}
\caption{Measured betatron tunes and nearby resonance lines. Approximate timing of major loss points are shown with triangle markers.}
\label{Figure22}
\end{center}
\end{figure}

Figure~\ref{Figure21} shows the measured fractional betatron tunes and beam loss throughout the acceleration cycle, and Fig.~\ref{Figure22} shows the measured tune excursion and nearby resonance lines, together with the approximate location of major beam loss points. 

\clearpage

\section{Stripping foil effects}

The eventual aim of the present collaboration is to explore the potential of FFAGs for high intensity operation. To move toward this, the effect of the stripping foil needs to be understood in detail. A first step in this direction is to measure the energy loss per turn caused by the stripping foil to produce an estimate of the foil thickness. For this, we undertook a campaign to measure the synchronous phase of the beam as a function of rf voltage. The procedure was as follows:

\begin{enumerate}
\item{With the RF switched off, we injected a short beam of $0.2\,\mu$s, less than one turn duration. For each revolution around the machine we observed the bunch charge monitor signal on an oscilloscope. We used this to measure the revolution frequency, then took this as the `set frequency' for the rf.}
\item{A value for the rf voltage was set and a longer beam of around $4\,\mu$s was injected with fixed rf frequency. We then compared the peaks in the bunch charge monitor signal with those of the rf signal (the set frequency) to determine the phase offset between the bunch and the rf signal. This step was repeated for various rf voltage levels.}
\item{Finally we examined the rf voltage and phase, which we compared with a simple one-dimensional model to determine the energy change per turn and hence foil thickness.}
\end{enumerate}

\begin{figure}
	\includegraphics[width=0.5\textwidth]{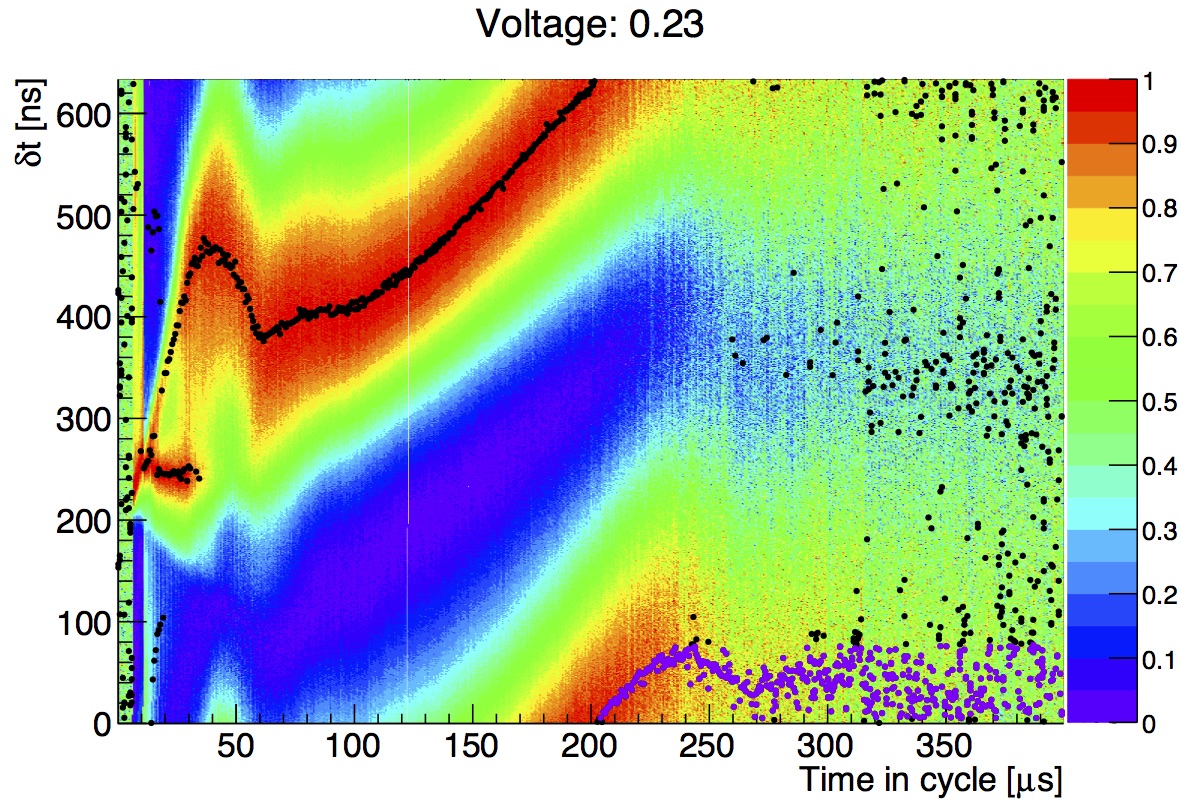}
 	\includegraphics[width=0.5\textwidth]{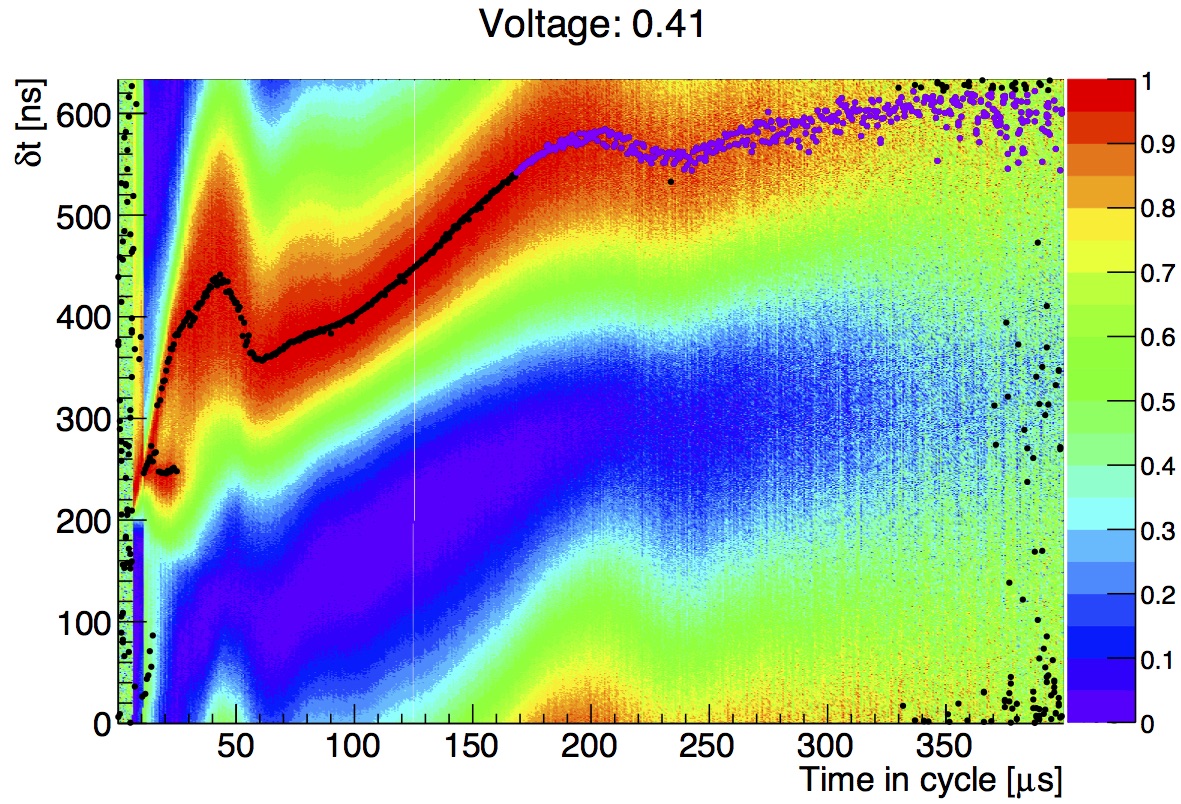}
    \includegraphics[width=0.5\textwidth]{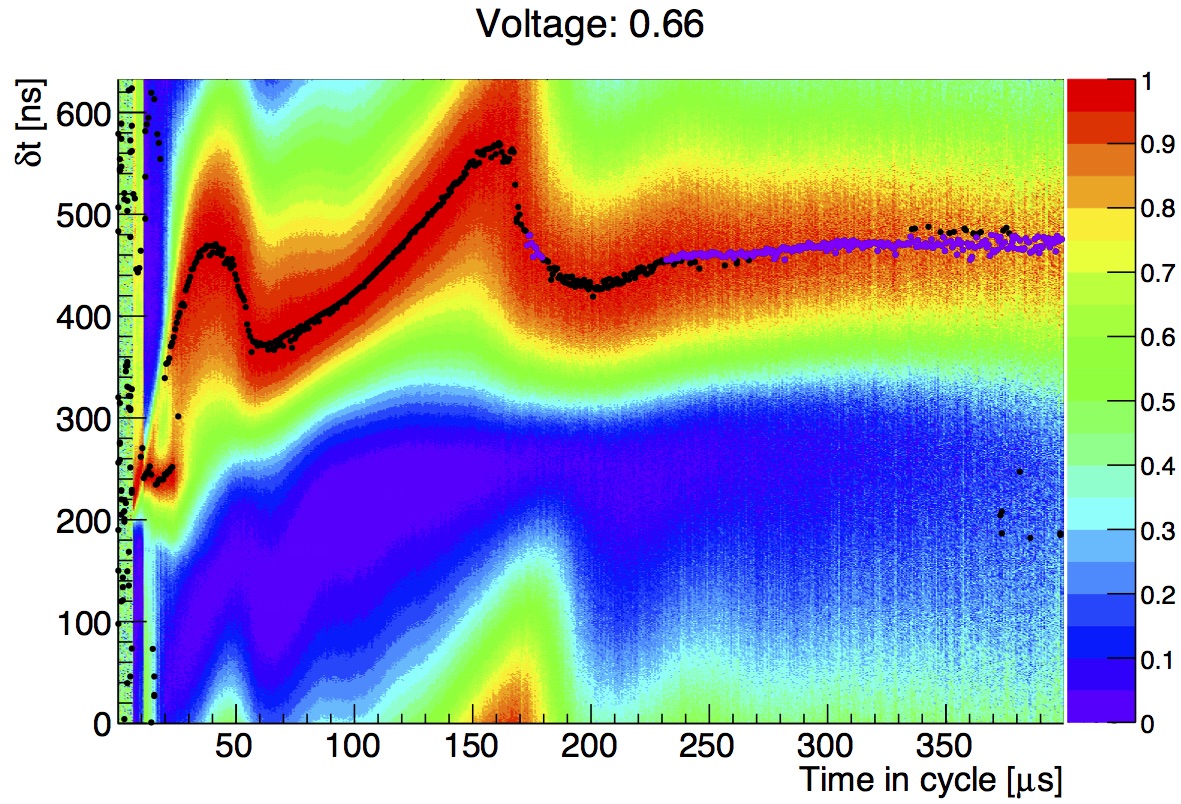}
	\includegraphics[width=0.5\textwidth]{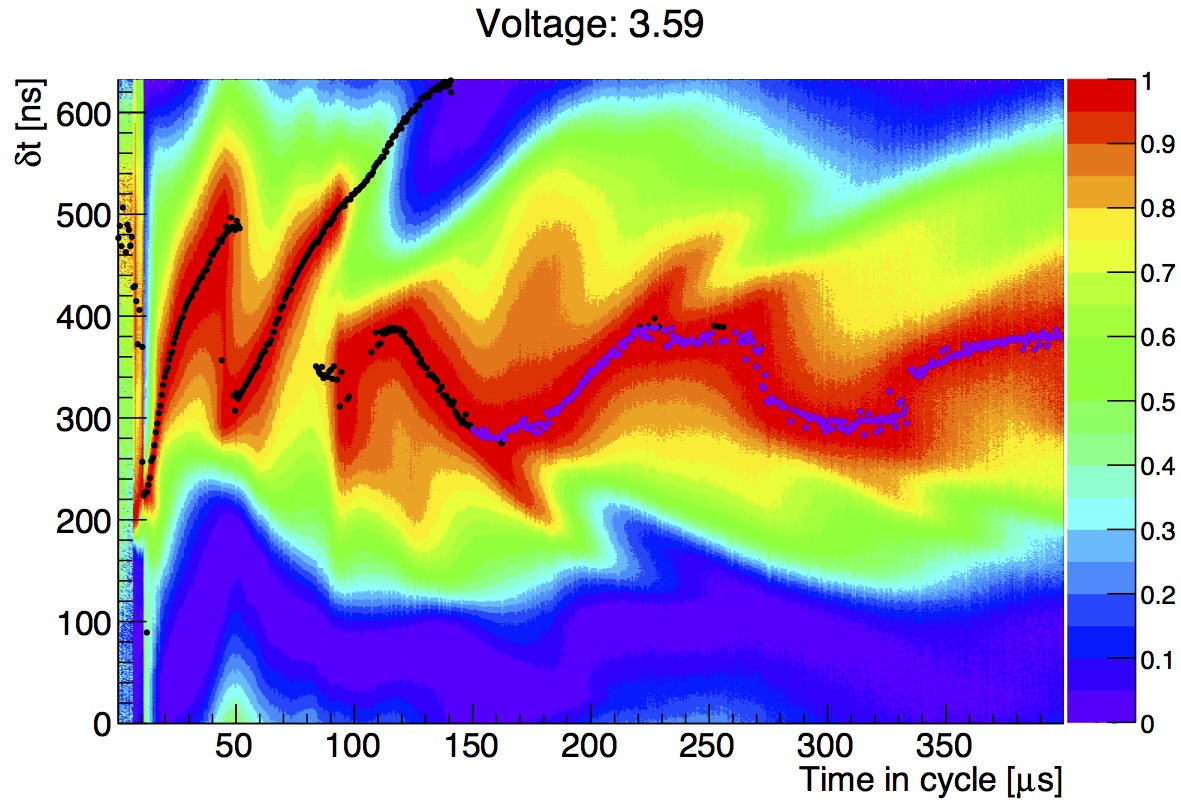}
	\caption{BPM signal for various rf voltages. $\delta t$ is the time relative to the most recent rf peak. The signal is normalized to lie within the range (0, 1) on each successive rf period. The points show the position of the measured peaks in the bunch charge monitor signal. Purple points were included in the rf phase analysis and black points were excluded after cuts applied to unbunched beam in the first 150 $\mu$s and filamented beam thereafter.}
	\label{fig:bm_signal_periodic}
\end{figure}

Sample longitudinal beam profiles are shown in  Fig.~\ref{fig:bm_signal_periodic} for a few lower voltage settings where the beam was not captured or barely captured. A higher voltage setting where the beam was well-captured is also shown. The captured beam rotates around the rf bucket. Nearly captured beam oscillates around the outside of the rf bucket before losing energy due to the foil and becoming lost. As the initial beam has a large time spread and small energy spread, this loss occurs with a periodicity of twice the synchrotron tune, leading to filaments becoming periodically lost from the beam. Note that the synchrotron tune is around 0.005 at injection.

It is possible to use the measured phase of the beam to constrain the energy loss due to the foil. In this situation, as the rf has fixed frequency, particles are captured in a stationary bucket. The voltage and phase of the stationary bucket provided by the rf cavity must match the energy lost on the foil for the bunch to be stable. We require
\begin{equation}
\label{eq:voltage_energy_loss}
cqV_{0}=dW/\sin(\phi_{s} + d\phi),
\end{equation}
where $V_{0}$ is the `read voltage', referring to a measurement of the voltage at the cavity, $c$ is a calibration constant that maps the read voltage to the potential difference that a particle in the beam undergoes, q is the unit charge. The calibration constant $d\phi$ represents the azimuthal offset of the rf cavity and bunch charge monitor, cable lengths and electronics delays, $dW$ is energy loss in the foil and $\phi_{s}$ is the measured phase offset.

\begin{figure}
		\includegraphics[width=1.0\textwidth]{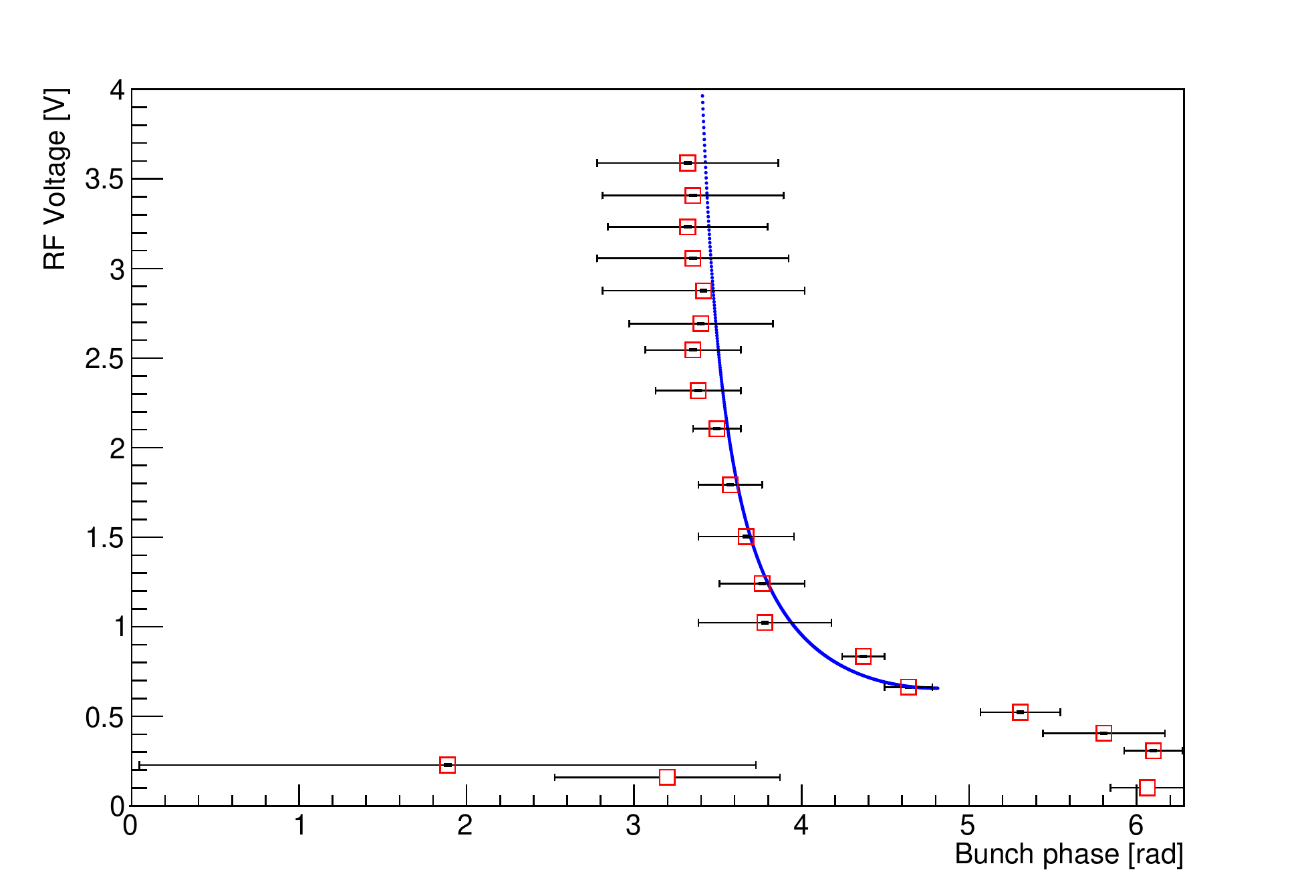}
	\caption{Measured synchronous phase compared to rf peak voltage. The fit curve shown is for a threshold voltage of 0.8353\,V.}
	\label{fig:rf_phase_to_voltage}
\end{figure}

The stable phase of the beam is estimated by seeking a region where the beam is stable following the initial capture. This is shown in Fig.~\ref{fig:rf_phase_to_voltage}. The width in $\delta$t of the stable region gives the estimated error.

We attempted to reconcile the measured values of $\delta$t and voltage with the model outlined in Eq. (\ref{eq:voltage_energy_loss}) by applying a chi-square test for goodness of fit to determine the $p$ value~\cite{Fisher92}. The results are shown in Fig. \ref{fig:rf_phase_to_voltage} and the relevant fit parameters are listed in Table \ref{tab:fitted_rf_voltage_and_phase}.

We were unable to reconcile the model with the theory for read voltages equal to or less than 0.664\,V, indicating that there is no measurable captured bunch in this situation. Based on this analysis, the phase of the bunch is not consistent with the model for voltages below 0.84\,V. The calibration factor from read voltage to actual rf voltage is 1000, so that the minimum rf voltage required to bunch the beam is measured to be at least 0.84\,kV, which we take as the measured energy lost in the foil.

\begin{table}
\begin{center}
\begin{tabular}[\linewidth]{|c||c|}
\hline
Lowest voltage & $p$ value \\
setting [V] &   \\
\hline
0.664  & 0.1 \\
0.840  & 0.83 \\
1.023  & 1.0 \\
1.241  &  0.99 \\
1.504  & 0.99 \\
\hline
\end{tabular}
\caption{Goodness of fit test result for different sets of minimum allowed voltage.}
\label{tab:fitted_rf_voltage_and_phase}
\end{center}
\end{table}

\subsection{Measured Thickness}
The measured energy lost in the foil is compared with the Bethe-Bloch model~\cite{PDG} and the GEANT4 QGSP energy loss model~\cite{GEANT4} to deduce a measured foil thickness. The results are shown in Table~\ref{tab:foilthickness}. The results obtained are consistent with the design foil thickness of 20\,$\mu$g/cm$^2$. A more precise result of the energy loss and thus foil thickness could be obtained by taking additional data points around the expected minimum voltage setting. 

\begin{table}
\begin{center}
\begin{tabular}[\linewidth]{|l||c||c|}
\hline
& Bethe Bloch   & GEANT4 QGSP model \\
\hline
Mean stopping power & 37.6 MeV cm$^2$ g & 34.0 MeV cm$^2$ g   \\
Fit & $22^{+5}_{-5}~\mu$g/cm$^2$    &  $25^{+6}_{-5}~\mu$g/cm$^2$ \\
\hline
\end{tabular}
\caption{Model stopping power and corresponding foil thickness for the energy loss value of 0.84\,kV, where the upper and lower error ranges correspond to energy loss of 1.023\,kV and 0.664\,kV respectively.}
\label{tab:foilthickness}
\end{center}
\end{table}

\section{\label{section7} Discussion}

\subsection{Different k in three azimuthal locations}

In the ideal machine all the orbits with different momenta should be similar so that the momentum compaction factor or $k$ value should be unique, independent of the measurement location. The experimental results in Section~\ref{Section3} show that the measured $k$ value in fact depends on the azimuthal location. This variation can be understood by considering how the local closed orbit (on which this measurement is based) depends on momentum.

The closed orbit at any point in the ring $r_i$ can be expressed as a sum of the ideal closed orbit and the distortion introduced by dipole kicks $\theta_j$. 

\begin{equation}
r_i = r_{i0} \left(\frac{p}{p_0}\right)^{\frac{1}{k+1}} + \sum_j R_{ij} \theta_j
\label{orbitwithcod}
\end{equation}

Note that the contribution of the high-order multipoles to the COD is ignored in Eq.~(\ref{orbitwithcod}). The validity of this approximation is discussed in Appendix B. The linear closed orbit response $R_{ij}$ at observation point $i$ caused by a kick at $j$, is given by  
\begin{equation}
R_{ij} = \frac{\sqrt{\beta_i \beta_j}}{2 \textrm{sin} \pi q_x} \textrm{cos}\left(|\psi_i - \psi_j| - \pi q_x\right)
\label{response}
\end{equation}
where $q_x$ is the horizontal tune, $|\psi_i - \psi_j|$ is the phase advance between elements and $\beta_{i,j}$ is the betatron function. Since $\beta_{i,j}$ grows linearly with radius in a scaling FFAG, $R_{ij}\propto (p/p_0)^{1/(k+1)}$.

In Section 3.1, the local closed orbit $r_i$ was used in place of the equivalent radius $R$ in Eq.~(\ref{Eqn4}) in order to calculate the momentum compaction factor $\alpha_p$. It is clear from Eq.~(\ref{orbitwithcod}) that the measured $\alpha_p$, and by implication the measured $k$, will vary with azimuthal location unless $\theta_j$ is independent of momentum. 

\subsection{Sources of COD}

Since the COD is measured at just three azimuthal locations, it is challenging to determine the error sources precisely. However, it is possible to test whether the predicted COD pattern produced by a suspected error source, for example stray field in the vicinity of the rf cavity, is consistent with measurements.

By taking the difference of the closed orbit data shown in Fig.~\ref{fig:r_k_vs_p}, the evolution of the COD amplitude and shape with momentum can be measured. In order to parameterise the latter, we introduce a shape parameter $\xi$ which is given by the ratio of the differences between the closed orbits measured at the three probes. 
\begin{equation}
\xi = \frac{r_1(p) - r_7(p)}{r_5(p)-r_1(p)}
\label{shape1}
\end{equation}
where the indices refer to the cell in which the probe is located. 

Eq.~(\ref{response}) may be used to express the shape parameter in terms of tune and error source location. Assuming symmetry, the phase difference is replaced with $2\pi q_x \Delta n_i/n$ where $n$ is the number of cells and $\Delta n_i$ is the number of cells between probe i and the error source. In the case of just a single error source (and ignoring the effect of higher-order multipoles), the shape parameter may be written
\begin{equation}
\xi = \frac{\textrm{cos}\left(2\pi q_x \Delta n_1/n - \pi q_x\right) - \textrm{cos}\left(2\pi q_x \Delta n_7/n - \pi q_x\right)}{\textrm{cos}\left(2\pi q_x \Delta n_5/n - \pi q_x\right) - \textrm{cos}\left(2\pi q_x \Delta n_1/n - \pi q_x\right)}
\label{shape2}
\end{equation}
The direct measurement of $\xi$ in Eq.~(\ref{shape1}) may be compared with the prediction given by Eq.~(\ref{shape2}) assuming the measured tune and some location for the error source. 

As can be seen in Fig.~\ref{Figure21}, the horizontal tune approaches integer during the acceleration cycle, increasing from about 3.65 at injection to about 3.85 at the maximum momentum. Assuming a single error source at the rf cavity, and using Eq.~(\ref{shape2}), it is found that  these tunes correspond to a decrease in $\xi$ from $3.05 \pm 0.02$ to $2.42 \pm 0.01$. From the closed orbit data,  and using Eq.~(\ref{shape1}), it is found that $\xi$ decreases from $2.8 \pm 0.2$ at injection to $0.93 \pm 0.03$ at the highest momentum.  The uncertainties were found by propagating estimated tune measurement and closed orbit measurement errors. Therefore, while it can be asserted that the COD at injection is dominated by a single error source at the rf cavity, the subsequent deviation of the shape parameter $\xi$ from the expected value may be the result of the tune approaching integer (see Appendix~\ref{appendixb}).

While stray fields may account for these additional errors, another possibility is that there is a variation in the field index from magnet to magnet. Similarly to Eq.~(\ref{Eqn6}), this local field index $k_i$ may be written
\begin{equation}
k_i = \frac{r_i}{B_i} \frac{dB_i}{dr_i}
\end{equation}
Here we consider $r_i$ and $B_i$ to be the closed orbit and magnetic field at the center of a hard-edge magnet. Rearranging as follows

\begin{equation}
\frac{\Delta B_i (k)}{B_i} = k_i \frac{\Delta r_i}{r_i}
\label{eqn:localk}
\end{equation}

The incremental change following $k_i \rightarrow k_i + \Delta k_i$ is then
\begin{equation}
\frac{\Delta B_i (\Delta k)}{B_i} = \Delta k_i \frac{\Delta r_i}{r_i}
\label{eqn:inc}
\end{equation}

It follows that the ratio of the resulting dipole kick $\theta_{\Delta k}$ and the bending angle $\theta_i$ in magnet $i$ is
\begin{equation}
\frac{\theta_{\Delta k}}{\theta_i} = \frac{\Delta B_i (\Delta k_i) }{B_i} = \Delta k_i  \frac{\Delta r_i}{r_i}.
\label{eqn:deltak}
\end{equation}

By increasing the number of points around the ring at which the closed orbit is measured, it should be possible to locate the additional error sources.

\subsection{Plans for future experiments and upgrades}

A number of parameters have not been measured with the current set of diagnostics. As mentioned in Section~\ref{Section4} we cannot at present measure the optics functions (ie. Twiss parameters) and emittance in this machine. Further diagnostics which would allow measurement of beam size, emittance and Twiss parameters are under investigation.

In the near future a number of steps will be undertaken toward the aim of machine optimization and high power operation. Various studies have been highlighted throughout this work, these include:

\begin{itemize}
\item{Calibrating the radially movable horizontal BPM for position readout.}
\item{Further correction of the closed orbit distortion with a stronger corrector field.}
\item{Optimization of dispersion in the injection line to match the ring.}
\item{Developing diagnostics and techniques for emittance and optics measurements.}
\item{Further measurement to refine the energy loss on the stripping foil.}
\end{itemize}

Another area which has not been addressed in the present work is optimization of the rf profile. It is clear in Fig.~\ref{Figure21} that there is a large amount of loss at injection during capture of the beam. Careful optimization of the rf bucket height, synchronous phase and frequency profile are expected to yield improvements in beam transmission. This work is planned for the near future.

However, the most significant future alteration to the machine will be the addition of a second RF cavity in the ring. This will increase the acceleration rate which will have a positive impact in many respects, however it may create further COD which will have to be corrected. The overall impact on the operation of the machine remains to be studied.

\section{\label{section8} Summary}

In this paper we have described a series of experiments and techniques to characterize the 150 MeV proton scaling Fixed Field Alternating Gradient (FFAG) accelerator at KURRI. We have identified key diagnostics and outlined their use in various measurements in this class of accelerator, introducing variations on existing accelerator diagnostics such as the radially movable triangle BPM.

To improve the beam quality of the FFAG and pursue high intensity operation, we must first understand the operation of the machine at low intensity. We have used the unique property of the FFAG, that is the shifting orbit position and momentum with rf frequency, together with simple radial probes and bunch charge monitors to determine the momentum compaction factor or $k$ value and periodic dispersion in FFAGs without the need for assumptions based on simulation.

We have shown how these same techniques can be used to determine the level and pattern of closed orbit distortion and shown the efficacy of the correction scheme. Further correction is desired in this machine and will be pursued in the near future. We have also developed methods to match the horizontal and vertical orbit position by reducing measured coherent oscillations, a technique which would be quick and accurate with an increased number of calibrated vertical and horizontal BPMs. We have also undertaken significant work toward understanding and matching the dispersion from the injection line to the ring, demonstrating control over the dispersion in the injection line and outlining various methods for measuring dispersion at different locations. 

We have reported a number of experimental results which would not be expected in a perfectly scaling FFAG, such as the variation of betatron tune with momentum. This information will help to optimize the working point of the FFAG for future operation. Finally, we have devised a beam-based method of measuring energy loss of the beam on the stripping foil, and deduced from this a measured foil thickness. While the accuracy of this measurement would be improved with more data, this is a technique which may also be of use in the wider accelerator community.

With this level of understanding, further experiments can now be undertaken to optimize the acceleration parameters, to ascertain the impact of future upgrades and to begin planning a route to increasing the intensity of the accelerator.

%\section{Acknowledgements}
%The authors would like to thank J. S. Berg and other members of the FFAG community for their insight and discussion of experiments and analysis.

\bibliographystyle{ptephy}
\bibliography{kurripaper-1.bib}

\appendix

\section{Equivalent momentum method}
\label{appendixa}

We discuss in detail how we set the equivalent momentum used for dispersion measurements. In order to set up an equivalent momentum at $+dp/p$, the current of the focusing magnet was changed by $dI_F/I_F=-dp/p$ and that of the defocusing magnet by $dI_D/I_D=-dp/p$. However, because the strength of the magnet is not necessarily proportional to the excitation current, the change in field strength of the focusing and defocusing magnets could be different. First we calculate the absolute change of field strength corresponding to a change of excitation current, which was estimated using TOSCA field calculation software. The measured B-H curve was supplied in the calculation. 

\begin{table}[!h]
\caption{Magnet strength with varying excitation current.}%%%Table caption goes here
\label{table:magneticfield}
\centering
\begin{tabular}{|c||c||c|}%%%The number of columns has to be defined here
\hline
Current factor & Power supply & Field change\\ %%%% Table body
 &						$I_{F}$/$I_{d}$ & $B_{F}$/$B_{D}$\\
\hline
-5\% & 773.30 / 1140 & -2.86\% / -2.75\%\\
-2\% & 797.72 / 1080 & -1.10\% / -0.81\%\\
0\% & 814.00 / 1012 & 0\% / 0\%\\
2\% & 830.28 / 940 & +0.87\% / +0.80\%\\
\hline
\end{tabular}
\end{table}%%%End of the table

From this we know that the strength of each magnet is not proportional to the excitation current, so we cannot simply change the excitation current of both magnets by the same amount. Instead, in order to adjust them correctly, the excitation current of the defocusing magnet was searched so that the vertical tune remained constant. We choose to fix the F field strength and vary D because the average radius is mainly determined by the F magnets. Since the vertical tune is particularly sensitive to the ratio of F and D magnet strength, finding a setting which produces the same vertical tune means that the D magnet field strength was changed by the same ratio as that of the F magnet. Thus we can be sure that the F/D ratio has been maintained while the equivalent momentum has been scaled.

\section{Validity of closed orbit formula}
\label{appendixb}

At any fixed momentum, the equation of motion in the horizontal plane in the presence of a dipole error field $\Delta B/B$ is given by 
\begin{equation}
x'' + \kappa_1 x = -Re\left[\sum_{n \ge 2} \frac{\kappa_n + i j_n}{n!}  (x + iy)^n  \right] - \frac{\Delta B}{B \rho}
\label{eqn:eom}
\end{equation}
where the differentiation is with respect to the longitudinal coordinate $s$, $\rho$ is the bending radius and $\kappa_n$ and $j_n$ are the normal and skew multipole terms, respectively. Note, Eq.~\eqref{eqn:eom} is defined with respect to the ideal closed orbit identified by $x=0$ when $\Delta B/B = 0$. 

The standard formula for COD (i.e. Eq. 19) is obtained when the first term on the RHS is neglected, i.e the nonlinearities are ignored. This solution is denoted $x_c(s)$.
 
In a scaling FFAG the magnetic field varies with $r^k$, where $r=r_0+x$ and $k$ is the field index, so the normal multipole components are
\[\kappa_n=\left.\frac1{B\rho}\frac{d^nB}{dx^n}\right|_{x=0}= \frac{k!}{\rho r_0^n\left(k-n\right)!}\,,\qquad j_n=0\,.\]
Considering only the sextupole component, Eq.~\eqref{eqn:eom} can be written as 
\[x'' + \kappa_1 x = -\frac{k(k - 1)}{2 \rho r_0^2}x^2 - \frac{\Delta B}{B \rho}\,.\]
This allows us to estimate, to a good approximation, the additional effect of the pseudo-kick from the sextupole term in any magnet via
\[\theta_\mathrm{sext} \approx \int_\mathrm{magnet} \frac{k(k-1)}{2\rho r_0^2}x^2 ds\,,\]
where $x$ is replaced by $x_c$.
Similarly, the kick angle from the dipole error field is measured by $\displaystyle\theta_x=\int\frac{\Delta B}{B\rho} ds $. The relative effect of the sextupole terms can then be ignored in comparison with dipole field error provided
\begin{equation}
 \max(\left|\theta_\mathrm{sext}\right|)\ll\left|\theta_x\right|\,.
\label{eqn:cond}
\end{equation}

A specific estimate may be made of the contribution of the sextupole term using values obtained for the ADSR-FFAG. At the injection momentum where $\left<r_i\right> = 4.6$\,m and $q_x = 3.65$, and assuming a smooth 
focusing betatron function $\beta_s= \left<r_i\right>/q_x$, the amplitude of the linear closed orbit response \[\left|R_{ij}\right|= \dfrac{\beta_s}{2\left|\sin\pi q_x \right|}= 0.7\,\mathrm{m}.\] Hence, 
$\theta_x= 40$\,mrad is required to produce the measured COD, $x=\pm 28$\,mm. Using the measured field index $k$ and noting $\rho\approx 1$\,m, the pseudo-kick from the sextupole
evaluated in any F magnet (arc length $\sim 0.8$\,m) is $\theta_\mathrm{sext} = 0.8$\,mrad (meeting the condition given by Eq.~\eqref{eqn:cond}). While it can be seen that the contribution of the nonlinear component
is negligible in this case, the effect will increase as the tune approaches integer (owing to the $1/\sin\pi q_x$ dependence of the linear closed orbit response).

\end{document}